%% file: main.tex
\pdfoutput=1
\documentclass[11pt]{article}

\usepackage[final]{acl}

\usepackage{times}
\usepackage{latexsym}
\usepackage{hyperref}
\usepackage{url}
\usepackage{xspace}
\usepackage{enumitem}
\usepackage{amsmath,amssymb,amsfonts}
\usepackage{mathtools}
\usepackage[linesnumbered,ruled]{algorithm2e}
\usepackage{algpseudocode} 
\usepackage{mathtools}

\usepackage[T1]{fontenc}
\usepackage[utf8]{inputenc}
\usepackage{microtype}
\usepackage{inconsolata}
\usepackage{graphicx}
\usepackage{colortbl}
\usepackage{threeparttable}
\usepackage{makecell}
\usepackage{multirow}
\usepackage[most]{tcolorbox}
\usepackage{adjustbox}
\usepackage{xcolor}
\usepackage{bbding}
\usepackage{booktabs}
\usepackage{makecell}

\definecolor{aigreen}{HTML}{a4ecad}
\definecolor{empty}{HTML}{a0d8ef}
\definecolor{checkgreen}{HTML}{4AA35A}
\definecolor{darksalmon}{rgb}{0.91, 0.59, 0.48}
\definecolor{background}{HTML}{FFFFF5}
\definecolor{edge}{HTML}{87C2B1}

\usepackage[most]{tcolorbox}
\newtcolorbox{mybox}{colback=background!35,
colframe=edge,
width=\columnwidth,
arc=1mm,
auto
outer
arc
}

\def\tool{{UniDebugger}\xspace}
\title{\tool: Hierarchical Multi-Agent Framework for Unified Software Debugging}

\author{
 \textbf{Cheryl Lee\textsuperscript{1}\textsuperscript{*}},
 \textbf{Chunqiu Steven Xia\textsuperscript{2}},
 \textbf{Longji Yang\textsuperscript{1}},
 \textbf{Jen-tse Huang\textsuperscript{1}},
 \\
 \textbf{Zhouruixing Zhu\textsuperscript{3}},
 \textbf{Lingming Zhang\textsuperscript{2}},
 \textbf{Michael R. Lyu\textsuperscript{1}} 
 \\
 \\
 \textsuperscript{1}The Chinese University of Hong Kong ~~
 \textsuperscript{2}University of Illinois Urbana-Champaign
 \\
 \textsuperscript{3}The Chinese University of Hong Kong, Shenzhen
 \\
\small{
    \textsuperscript{1}\textsuperscript{*}\href{mailto:cheryllee@link.cuhk.edu.hk}{cheryllee@link.cuhk.edu.hk}~~\textsuperscript{2}chunqiu2@illinois.edu}~~
\textsuperscript{6}lingming@illinois.edu~~ \textsuperscript{7}lyu@cse.cuhk.edu.hk
}

\newcommand{\locator}{\textit{Locator}\xspace}
\newcommand{\fixer}{\textit{Fixer}\xspace}
\newcommand{\fixerpro}{\textit{FixerPro}\xspace}
\newcommand{\summarizer}{\textit{Summarizer}\xspace}
\newcommand{\slicer}{\textit{Slicer}\xspace}
\newcommand{\repofocus}{\textit{RepoFocus}\xspace}
\newcommand{\helper}{\textit{Helper}\xspace}

\newcommand{\llm}{LLM\xspace}
\newcommand{\apr}{APR\xspace}
\newcommand{\fl}{FL\xspace}
\newcommand{\rag}{RAG\xspace}
\newcommand{\nmt}{NMT\xspace}
\newcommand{\astree}{AST\xspace}
\newcommand{\aprfull}{Automated Program Repair\xspace}
\newcommand{\flfull}{Fault Localization\xspace}

\newcommand{\flaw}{Codeflaws\xspace}
\newcommand{\quix}{QuixBugs\xspace}
\newcommand{\cond}{ConDefects\xspace}
\newcommand{\dfj}{Defects4J\xspace}

\newcommand{\alpharepair}{AlphaRepair\xspace}
\newcommand{\rewardrepair}{RewardRepair\xspace}
\newcommand{\chatrepair}{ChatRepair\xspace}
\newcommand{\angelix}{Angelix\xspace}
\newcommand{\prophet}{Prophet\xspace}
\newcommand{\spr}{SPR\xspace}
\newcommand{\semfix}{Semfix\xspace}
\newcommand{\genprog}{GenProg\xspace}
\newcommand{\coconut}{CoCoNuT\xspace}
\newcommand{\cure}{CURE\xspace}
\newcommand{\tbar}{Tbar\xspace}
\newcommand{\recoder}{Recoder\xspace}
\newcommand{\repilot}{Repilot\xspace}

\newcommand{\llama}{LLaMA2\xspace}
\newcommand{\codellama}{CodeLlama\xspace}
\newcommand{\deepseek}{DeepSeekCoderV2\xspace}
\newcommand{\gemini}{gemini-1.5-flash\xspace}
\newcommand{\claude}{claude-3.5-sonnet\xspace}
\newcommand{\chatgpt}{gpt-3.5-turbo-ca\xspace}
\newcommand{\gpt}{gpt-4o\xspace}

\newcommand{\sota}{SoTA\xspace}

\begin{document}
\maketitle

\input{sections/0_abstract}
\input{sections/1_introduction}

\input{sections/2_related}

\input{sections/3_approach}

\input{sections/4_experiments}

\input{sections/5_results}

\input{sections/6_ablation}
\input{sections/7_conclusion}

\input{sections/8_Limitation}
\input{sections/Acknowledgments}

\bibliography{custom}
\appendix
\onecolumn
\input{sections/appendix}

\end{document}

%% file: sections/0_abstract.tex
\begin{abstract}
Software debugging is a time-consuming endeavor involving a series of steps, such as fault localization and patch generation, each requiring thorough analysis and a deep understanding of the underlying logic.
While large language models (\llm{s}) demonstrate promising potential in coding tasks, their performance in debugging remains limited.
Current \llm-based methods often focus on isolated steps and struggle with complex bugs.
%
In this paper, we propose the first end-to-end framework, \textbf{\tool}, for unified debugging through multi-agent synergy.
It mimics the entire cognitive processes of developers, with each agent specialized as a particular component of this process rather than mirroring the actions of an independent expert as in previous multi-agent systems.
Agents are coordinated through a three-level design, following a cognitive model of debugging, allowing adaptive handling of bugs with varying complexities.
Experiments on extensive benchmarks demonstrate that \tool significantly outperforms state-of-the-art repair methods, fixing 1.25$\times$ to 2.56$\times$ bugs on the repo-level benchmark, \dfj.
This performance is achieved without requiring ground-truth root-cause code statements, unlike the baselines.
Our source code is available on an anonymous link: \url{https://github.com/BEbillionaireUSD/UniDebugger}.

\end{abstract}

%% file: sections/1_introduction.tex
\section{Introduction}
\begin{figure}[t]
    \centering
    \vspace{-0.1in}
        {\includegraphics[width=0.85\linewidth]{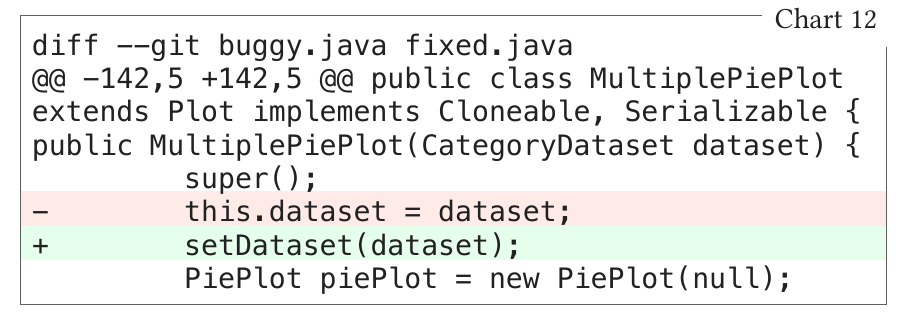}}
    \vspace{-0.1in}
    \caption{An example of the ``diff'' patch.}
    \vspace{-0.2in}
    \label{fig:patch-example}
\end{figure}

Debugging is a crucial process of identifying, analyzing, and rectifying bugs in software.
Significant advancements~\citep{LLMAO, AlphaRepair, Repilot, ChatRepair} have been achieved in addressing bugs with the boost of \llm{s}—they typically propose a prompting framework and query an \llm to automate an isolated phase in test-driven debugging, generally \flfull (\fl) or \aprfull (\apr).
\fl attempts to identify suspicious code statements, while \apr provides patches or fixed code snippets.
A typical ``diff'' patch records the textual differences between two source code files, as shown in Figure~\ref{fig:patch-example}. 

\begin{figure*}[t]
    \centering
    \vspace{-0.1in}
        {\includegraphics[width=0.9\linewidth]{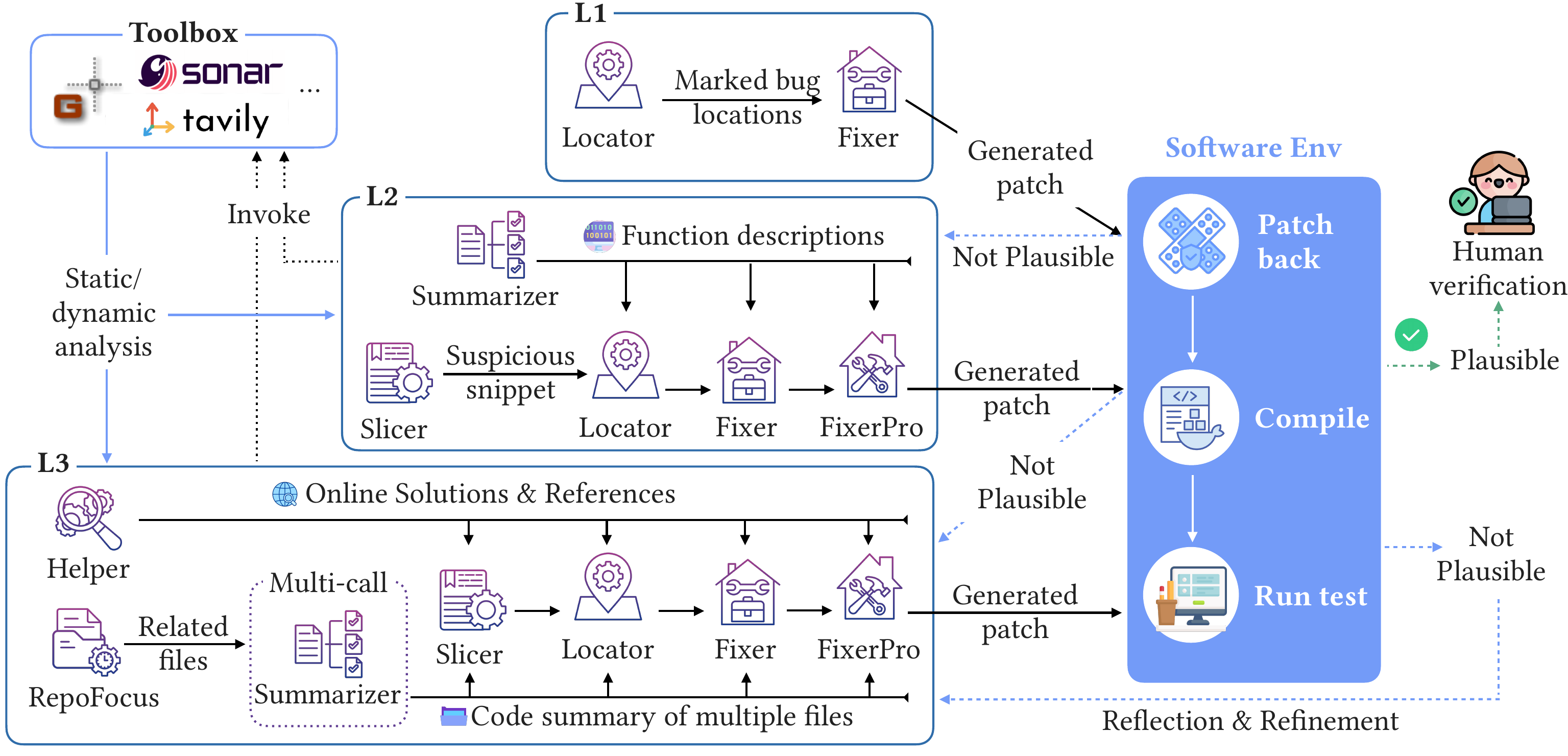}}
    \vspace{-0.1in}
    \caption{Overview of \tool. It starts with the simple \textit{L1} repair. If no plausible patch is generated, the \textit{L2} repair is triggered, and so is \textit{L3}. Agents on the same level can communicate with others.}
    \vspace{-0.2in}
    \label{fig:unidebugger}
\end{figure*}

While \llm{s} demonstrate great potential in addressing individual debugging tasks for basic bugs, previous APR studies cannot deliver a satisfying end-to-end solution for the entire task.
To bridge this gap, we propose the first multi-agent framework, \textbf{\tool}, for developer-side debugging, which can be seamlessly integrated into the CI/CD pipeline.
A key challenge lies in how to coordinate multiple agents. 
Existing multi-agent frameworks for complex problem-solving adopt a horizontal collaboration paradigm where each agent acts as an independent expert, mirroring human team dynamics through bidirectional discussions and mesh-like communication (e.g., ~\cite{MetaGPT, MapCoder, ChatDev}).
However, this design introduces critical limitations for debugging: (1) The inherent redundancy of peer-to-peer negotiations conflicts with debugging’s logical, incremental nature, wasting computational resources on trivial bugs; (2) The lack of structured problem decomposition leads to suboptimal resource allocation across bugs of varying complexity.

In response, we uniquely structure \tool as a hierarchical multi-agent coordination paradigm grounded in cognitive debugging theory, as shown in Figure~\ref{fig:unidebugger}.
\textit{Our key innovation lies in redefining agents as functional components of a unified cognitive process rather than autonomous experts.}
Inspired by Hale and Haworth’s model of structural learning~\citep{Hale}, which argues that developers employ a multi-level goal-orientated mechanism during debugging~\citep{HaleEvaluation}, \tool implements a three-level mechanism. 
The initial level only provides quick and straightforward solutions, and if that fails, higher levels of repairs are triggered for complex bugs, entailing more cognitive activities that involve deeper analysis, tool invocation, and external information ingestion.
Unlike mesh-based teamwork, where members with diverse perspectives and backgrounds engage in bidirectional discussion and negotiation to share knowledge and solve tasks collaboratively, this cognitive process is unidirectional and coherent, with each stage accumulating knowledge incrementally and building on the last.
By seamlessly aligning with debugging, this paradigm serves as the basis of our design for multi-agent synergy.

Our framework encompasses seven agents, each specialized in a distinct cognitive state: 1) \textbf{\helper}: retrieves and synthesizes debugging solutions through online research; 2) \textbf{\repofocus}: analyzes dependencies and identifies bug-related code files; 3) \textbf{\summarizer}: generates code summaries; 4) \textbf{\slicer}: isolates a code segment (typically tens to hundreds of lines) likely to cause an error; 5) \textbf{\locator}: marks the specific root-cause code lines; 6) \textbf{\fixer}: generates patches or fixed code snippets; 7) \textbf{\fixerpro}: generates an optimized patch upon the testing results of the patch generated by \textit{\fixer} along with a detailed analysis report.

Particularly, \tool first isolates the code file most likely causing the error through unit testing.
On level one (\textit{L1}), only \locator and \fixer are initiated for simple bugs.
If the generated patch is not plausible (i.e., pass all test cases), \summarizer and \slicer are triggered with the bug-located code file on level two (\textit{L2}). Then \locator, \fixer, and \fixerpro are called sequentially with access to the responses from \summarizer and \slicer.
If it still fails, \tool undertakes a deeper inspection and thus activates all agents (i.e., level three, \textit{L3}), where \helper searches for solutions online to guide all the other agents, and \repofocus examines the entire program to provide a list of bug-related code files, followed by the other agents working in line. 
\slicer, \locator, and \fixer can invoke traditional code analysis tools on the last two levels to collect static and dynamic analysis information.

We evaluate \tool on four benchmarks with bug-fix pairs across three programming languages, including real-world software (\dfj~\citealp{Defects4J}) and competition programs (\quix~\citealp{QuixBugs}, \flaw~\citealp{Codeflaws}, \cond~\citealp{ConDefects}).
On \dfj, \tool achieves a new state-of-the-art (\sota) by correctly fixing 197 bugs with 286 plausible fixes.
Our method does not require prior \fl while still outperforming strong baselines like \chatrepair, which uses ground-truth root causes with 10x sampling times.
Plus, \tool fixes all bugs in \quix and generates 2.2$\times$ more plausible fixes on \flaw.
Our ablation study shows that \tool consistently makes significant improvements across various \llm backbones. 


%% file: sections/2_related.tex
\section{Related Work}
\noindent \textbf{Software Debugging:}
Test-driven debugging has gained popularity since testing is an essential and prevalent part of practical CI/CD, and this methodology typically contains two core steps~\cite{UnifiedSurvey}: \fl and \apr.
Many studies aim to automate either step.
Traditional \fl are typically spectrum-based~\citep{SpectrumFL, Ochiai, SpectrumFL2011} or mutation-based~\citep{MutantFL, TraPT, InjectFL}.
Learning-based \fl learns program behaviors from rich data sources~\citep{DeepRL4FL, FixLocator, GRACE, DeepFL} via multiple types of neural networks. Recently, LLMAO~\citep{LLMAO} proposes to utilize \llm{s} for test-free \fl.
\apr research either searches for suitable solutions from possible patches~\citep{genProg1, genProg2, geneticAPR} or directly generates patches by representing the generation as an explicit specification inference~\citep{Angelix, SemFix, Tbar}.
Learning-based studies (\citep{CoCoNuT, CURE, RewardRepair}) translate faulty code to correct code using neural machine translation (\nmt).
On top of directly prompting \llm{s}~\citep{LLMAPREra}, recent studies have explored prompt engineering~\citep{ChatRepair, AlphaRepair} or combining code synthesis~\citep{Repilot} for \apr.
In addition to developer-oriented debugging, techniques for addressing user-oriented issues receive great attention~\cite{Agentless, SWE-Bench}.
These methods responsively update software based on problems discovered by users.
This paper focuses on the CI phase, reducing the burden on developers responsible for proactively detecting and resolving bugs to prevent errors from entering production.


\noindent \textbf{\llm-based Multi-Agent (\llm-MA):}
Various \llm{s} have been developed for code synthesis~\citep{Codex, DeepSeekCoder, CodeLlama}, in addition to general-purpose \llm{s}~\citep{GPT3-5, GPT4, Gemini}. They have shown potential in solving coding-related tasks, including program repair~\citep{LLMAPREra, LLMCSurvey, DebugBench}.
The inspiring capabilities of the single \llm-based agent boost the development of multi-agent frameworks. Recent research has demonstrated promising results in utilizing \llm-MA for complex problem-solving, including software engineering, science, and other society-simulating activities.
For software engineering, \llm-MA studies~\citep{ChatDev, MetaGPT, Self-collaboration, AgentCoder, MapCoder} usually emulate real-world roles (e.g., product managers, programmers, and testers) and collaborate through communication. \citet{MetaGPT, Self-collaboration} adopt a shared information pool to reduce overhead.

%% file: sections/3_approach.tex
\section{\tool}
\tool is an end-to-end framework for unified debugging through \llm-based multi-agent synergy.
It comprises seven agents, each specialized as a state in Hale and Haworth’s cognitive debugging model with an explicit goal instead of being a task-oriented, individual entity.
\tool operates on a three-level architecture, initializing different levels of repair involving different agents, and agents can adaptively invoke tools in a given toolbox.
The communication among agents on the same level follows an assembly-line thought process rather than the mesh interactions typical of teams.
%
This section introduces the profiles of agents, their interactions with external environments, and the hierarchical organization.
Prompt details are displayed in~\ref{sec:app:prompt}.

\subsection{Profiles of Agents}
All agents in \tool share a one-shot structure of the system prompt, which defines their roles, skills, actions, objectives, and constraints, followed by a manually crafted example to illustrate the desired response format. 
Moreover, all agents have access to certain meta-information known as \textit{bug metadata}, including the bug-located code file, failing test cases, reported errors during compilation and testing, and program requirements described in natural language.
The response of each agent comprises two elements: the answer fulfilling its goal and an explanation of its thinking.
Inspired by a software engineering principle, rubber duck debugging~\citep{RubberDuck}, where developers articulate their expectations versus actual implementations to identify the gap, we request each agent to monitor key program variables and explain how it guides the answer.

\begin{figure}[t]
    \centering
    \vspace{-0.1in}
    {\includegraphics[width=0.99\linewidth]{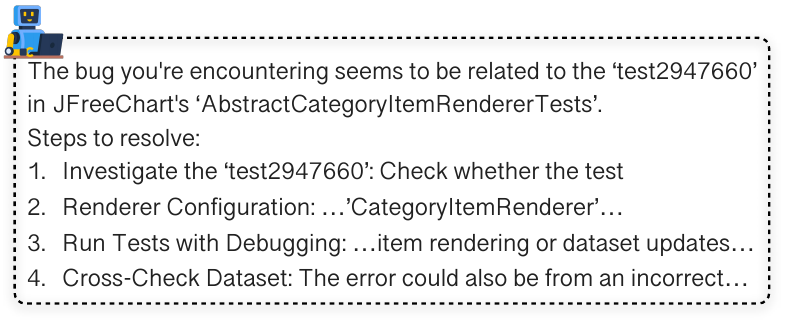}}
    \vspace{-0.3in}
    \caption{After retrieving similar solutions, \helper eventually responds with an executable debug guide.}
    \vspace{-0.2in}
    \label{fig:helper}
\end{figure}

\noindent \textbf{Helper.}
The goal of \helper is to provide references via retrieve-augmented generation (\rag), inspired by the fact that developers commonly utilize web search engines (e.g., Google) to enhance productivity~\citep{DevSearch}.
By analyzing the bug metadata, \helper generates a short query and invokes an external search engine to retrieve the best-matching solutions.
Then, it integrates the retrieval results and generates reference solutions that are customized for the context of the buggy code (e.g., variable naming and function structure).
These are internal processes hidden from the other agents, and only the final response—a reference solution—can be read only on \textit{L3}.
Figure~\ref{fig:helper} provides an example of the response from \helper.

\noindent \textbf{RepoFocus.}
Comprehending cross-file dependencies is crucial to debugging large software, and real-world software is often large and complex (corresponding to \textit{L3} repair).
However, inputting all of the code in a program may overwhelm an \llm, resulting in slow responses with significant resource consumption or incorrect results. 
Thus, \repofocus provides a list of bug-related code files that need further examination by analyzing the bug metadata and the file structure of the program.

\noindent \textbf{Summarizer.}
Code summarization aims to generate concise, semantically meaningful summaries that accurately describe software functions. Unlike high-level code analysis, these natural language summaries better align with the training objectives of \llm{s}.
\summarizer is triggered on \textit{L2} and \textit{L3}.
On \textit{L2}, it summarizes the buggy code file.
Though the window size of advanced \llm{s} can handle most single code files, a too-long prompt may exacerbate the illusion of \llm{s}. 
Thus, \summarizer handles overly long code here, in conjunction with \slicer, which narrows down a suspicious segment from the buggy code.  
On \textit{L3}, \summarizer runs once on every bug-related file identified by \repofocus.
The other agents do not need to read every code line in the program while still knowing the core contents through these summaries.

\noindent \textbf{Slicer.}
\slicer narrows down the inspection scope by slicing out a small suspicious segment (typically tens of code lines) from the bug-located code file.
We extract the beginning and end lines from the initial output to locate the segment, ensuring the final output is directly sliced from the original code to prevent code tampering caused by \llm hallucinations.
\slicer is also launched for large software on the last two levels, where it can invoke static or dynamic analysis tools.

\noindent \textbf{Locator.}
\locator is responsible for marking code lines with a comment ``// buggy line'' or ``// missing line'' to indicate faulty or missing statements.
Similar to \slicer, we directly annotate the original code through contextual string matching.
\locator only leverages the bug metadata on \textit{L1}, where on \textit{L2}, it can access the dynamic analysis reports. 
If the program is large enough so that \slicer and \summarizer are invoked, \locator receives the suspicious code segment (generated by \slicer) to replace the original complete code, along with the buggy code summary (generated by \summarizer).

\noindent \textbf{Fixer.}
\fixer is prompted to generate a "diff", as shown in Figure~\ref{fig:patch-example} in Introduction.
It receives code marked by \locator and other bug metadata.
To maintain consistency between \locator and \fixer, \fixer first assesses whether the marked lines should be modified and then describes the modification. This also enables \fixer to correct possible errors made by \locator.
On \textit{L2} and \textit{L3}, \fixer also has access to static/dynamic analysis and auxiliary information generated by upstream agents.

\begin{figure*}[t]
    \centering
    \vspace{-0.1in}
        {\includegraphics[width=0.99\linewidth]{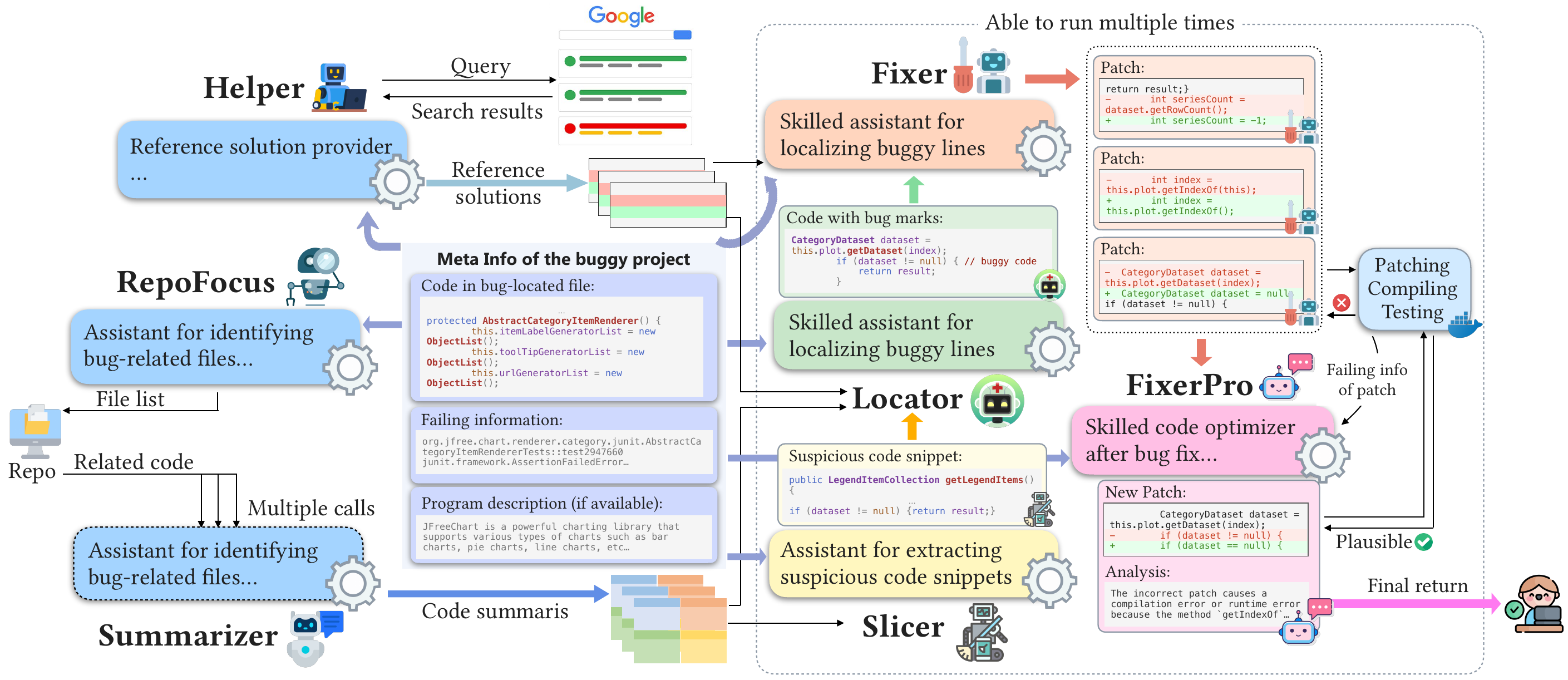}}
    \vspace{-0.1in}
    \caption{\textit{L3} triggers all the seven agents to generate plausible patches.}
    \vspace{-0.2in}
    \label{fig:level3}
\end{figure*}

\noindent \textbf{FixerPro.}
\fixerpro extends and complements \fixer by generating an optimized patch and repair analysis.
Inspired by code review, in which one or more developers check a program to ensure software quality, we request \fixerpro to evaluate the performance and potential vulnerability of the plausible fix generated by \fixer.
Then, \fixerpro provides suggestions for refactoring the patch to optimize simplicity and maintainability.
If \fixer fails, \fixerpro generates a new patch while analyzing the failing reasons.

\subsection{External Interactions}
\noindent \textbf{Plausibility Feedback.}
We conduct repairs by modifying the buggy code rather than directly applying the generated patches because they usually make mistakes in the format, especially the indices of code lines, as \llm{s} often struggle with counting.
Unlike prior studies, such as ~\cite{Agentless}, using a Search/Replace method for code editing, our post-adjustment methodology more readily aligns with LLM training databases as the outcomes of extensive code edits are stored in a "diff" format. 
Our modification is rule-based by matching the invariant context between the patch and the buggy code.
Afterward, we compile and run tests on the fixed program. 
The patch will be sent to developers for manual \textit{correctness} verification if passing all test cases.
The results of plausibility testing are returned to \tool as feedback for further processing, such as initializing a higher level of repair or conducting self-reflection.

\noindent \textbf{Tool Usage.}
The toolbox of \tool currently contains static analysis tools, dynamic analysis tools, and a search engine optimized for \llm{s} and \rag.
Static analysis provides static errors and warnings, as well as Abstract Syntax Tree (\astree), a tree representation of the abstract syntactic structure of source code.
We mainly consider the coverage of failing test cases in dynamic analysis since it is commonly assumed that code statements not executed during testing are less likely to cause the failures~\cite {FLSurvey}.
Note that the results of tool invocations will be stored for the requests of downstream agents to avoid duplicated invocations.


\subsection{Hierarchical Coordination}
Our main principle of coordination is that problems of different complexities warrant cognitive activities of different intensities.
That is, upon the failure of a straightforward solution, a higher-level goal will be triggered with more information and thinking.
Following Hale and Haworth’s cognitive debugging model, we define three workflows for the three levels of repair, respectively.

The first level identifies and fixes obvious logic errors in code, so we assume that agents can easily fix the bug. 
Thus, \textit{L1} only contains \locator and \fixer with a simple communication flow, where \fixer generates a patch based on the fault localization done by \locator.
If the fix is not plausible, then \textit{L2} is triggered, assuming that the fix can be achieved by only inspecting a single file, which is often applied in previous \fl and \apr studies~\citep{Assumptions}.
\textit{L2} involves five agents: \slicer, \summarizer, \locator, \fixer, and \fixerpro. 
\slicer isolates a suspicious segment, and \summarizer generates a code summary simultaneously, both from the bug-located file. Afterward, \locator, \fixer, and \fixerpro work in sequence based on the suspicious segment, and all have access to the code summary.
If it still fails, we turn to \textit{L3}, which believes that the bug is very complex, so repairing it requires understanding cross-file dependencies and external information.
\textit{L3} activates all the seven agents.
First, \helper and \repofocus are initialized at the same time, and \helper provides references for all other agents.
Then, \summarizer runs multiple times following the file list identified by \repofocus. These code summaries are shared by the remaining four agents, which work one after another subsequently with the same workflow as \textit{L2}.
Figure~\ref{fig:level3} presents an example conversation among agents on \textit{L3}.

Upon no plausible patch, \tool gradually requests agents on \textit{L3} to reflect and refine its answer based on the plausibility feedback in a reversed order, as downstream agents are more error-prone as errors may accumulate during the debugging process.
Specifically, we first ask \fixerpro to refine its response. If the fix fails again, we then request \fixer, leading to input changes of \fixerpro, so it is re-sampled the second time.
Eventually, plausible patches are returned to developers for manual verification.
If no plausible patch is produced, \tool will display the analysis report written by \fixerpro and take another run until reaching a resource threshold.

%% file: sections/4_experiments.tex
\input{tables/1_overall}

\section{Experiments}
\subsection{Experimental Setup}
\noindent \textbf{Benchmarks.}
We evaluate \tool on four benchmarks featuring bug-fix pairs, including competition programs (\flaw~\citealp{Codeflaws}, \quix~\citealp{QuixBugs}, \cond~\citealp{ConDefects}) and real-world projects (\dfj~\citealp{Defects4J}).
\flaw contains 3902 faulty C programs.
\quix includes 40 faulty programs available in both Java and Python.
\cond recently collected to address data leakage concerns in \llm{s}, consists of 1,254 Java and 1,625 Python faulty programs.
\dfj, a widely used benchmark from 15 real-world Java projects, features bugs across two versions: version 1.2 with 391 active bugs and version 2.0 adding an additional 415 active bugs, totaling 806. We will report the total number of fixes across these versions.

\noindent \textbf{Baselines.}
We compare \tool against 14 \apr baselines, including:
\begin{itemize}[topsep=0pt, itemsep=0pt, parsep=0pt]
    \item Traditional: \angelix~\citep{Angelix}, \prophet~\citep{Prophet}, \spr~\citep{SPR}, and CVC4~\citep{CVC4}.
    \item Genetic programming-based: \semfix~\citep{SemFix} and \genprog~\citep{genProg1, genProg2}.
    \item \nmt-based: \coconut~\citep{CoCoNuT}, \cure~\citep{CURE}, and \rewardrepair~\citep{RewardRepair}.
    \item Domain knowledge-driven: \tbar~\citep{Tbar} and Recoder~\citep{Recoder}.
    \item \llm-based (\sota):  \alpharepair~\citep{AlphaRepair}, \repilot~\citep{Repilot}, and \chatrepair~\citep{ChatRepair}.
\end{itemize}

Note that different \apr baselines adopted diverse prior fault locations. We use \textit{realistic} to represent traditional \fl and \textit{perfect} to denote ground-truth \fl. We report the results provided in their original papers and follow-up survey~\cite{Overfitting, QuixBugsSurvey, LLMAPREra}, following previous studies~\cite{LLMAPREra, ChatRepair, CoCoNuT}.

\noindent \textbf{\llm Backbones.}
We apply \tool to seven \llm{s}, including four general-purpose models (\gemini~\citealp{Gemini}, \chatgpt~\citealp{GPT3-5}, \gpt~\citealp{GPT4}, and \claude~\citealp{Claude}) and three open-source code \llm{s} (\deepseek~\citealp{DeepSeekCoder}, \codellama-34b~\citealp{CodeLlama}, \llama-70b~\citealp{LLaMA2}).
Naturally, these \llm{s} also serve as baselines for comparison, where we apply the original chain-of-thought (CoT) prompting method~\cite{CoT}.
The default backbone of \tool is \gpt.

\noindent \textbf{Metrics.}
We utilize the \apr metrics, specifically focusing on the number of bugs that are plausibly or correctly fixed. Given the extensive size of \cond, we randomize samples from plausible fixes to verify their correctness.
Thus, we also employ the metric known as \textit{correctness rate}—defined as the ratio of correct fixes to plausibly fixed bugs—to assess the effectiveness of \tool.

\noindent \textbf{Implementation.}
For single-file competition programs, only agents on \textit{L2} and \textit{L1} need to be initialized, named as \textbf{\textit{\tool-Lite}}.
The maximum number of attempts is set to 5 per bug for Lite and 20 for Full. Each query has a timeout limit of 1 sec.
Previous studies typically sample hundreds to thousands of times. For instance, \coconut samples up to 50,000 patches per bug~\cite{CoCoNuT}, while \chatrepair samples 100-200 times~\cite{ChatRepair}.
Our approach is significantly more cost-effective than baselines.
Furthermore, we alter the default web search engine (Tavily~\citealp{Tavily}) to a local one during evaluation to prevent retrieving ground-truth solutions online.
The local database consists of the training dataset of CoCoNut~\citep{CoCoNuT} with JavaScript programs removed. Since \coconut was also evaluated on \dfj and \cond, the risk of data leakage is minimal.
The static and dynamic analysis is supported by our written scripts and open-source plugins (\citealp{SonarQube} and \citealp{GZoltar}).

%% file: tables/1_overall.tex
\begin{table*}[htb]\centering
    \caption{Comparison with baselines. \textit{\#Corr} and \textit{\#Plau} represent the number of bugs correctly and plausibly patched, respectively. The green cells indicate the best results. The blue cells indicate the results are obtained from sampled data, while other results are obtained on the whole dataset.}
        \begin{threeparttable}
        \begin{adjustbox}{max width=\columnwidth*2}
        \begin{tabular}{*{11}{c}}
        \toprule
        \multirow{2}*{\textbf{Tools}} & \multirow{2}*{\makecell{\textbf{Sampling} \\ \textbf{Times}}}
        & \multicolumn{2}{c}{\textbf{\dfj-Java}} & \multicolumn{2}{c}{\textbf{\flaw-C}} & \multicolumn{2}{c}{\textbf{\quix-Java}} & \multicolumn{2}{c}{\textbf{\quix-Python}} & \multirow{2}*{\textbf{Note}} \\
        
        \cmidrule(lr){3-4}\cmidrule(lr){5-6}\cmidrule(lr){7-8}\cmidrule(lr){9-10}
        
        && \textit{\#Corr} & \textit{\#Plau} & \textit{\#Corr} & \textit{\#Plau} & \textit{\#Corr} & \textit{\#Plau} & \textit{\#Corr} & \textit{\#Plau} \\
        
        \cmidrule{1-11}\morecmidrules\cmidrule{1-11}
        
        \angelix & 1,000& - & - & 318 & 591 & - & - & - & - & \multirow{6}*{\rotatebox{90}{Realistic \fl}} \\
        
        \prophet & 1,000 & - & - & 310 & 839 & - & - & - & - \\
        
        \spr & 1,000 & - & - & 283 & 783 & - & - & - & - \\
        
        CVC4 & - & - & - & \cellcolor{empty}15$^\dagger$ & \cellcolor{empty}91$^\dagger$ & - & - & - & - \\

        \semfix & 1,000 & 25 & 68 & \cellcolor{empty}38$^\dagger$ & \cellcolor{empty}56$^\dagger$ & - & - & - & - \\

        \recoder & 100 & 72 & 140 & - & - & 17 & 17 & - & - \\ 
        
        \midrule 
    
        \genprog & 1000 & 5 & 20 & 255-369 & 1423 & 1 & 4 & - & - & \multirow{7}*{\rotatebox{90}{Perfect \fl}} \\

        \coconut & 20,000 & 44$^\star$ & 85$^\star$ & 423 & 716 & 13 & 20 & 19 & 21 \\

        \cure & - & 57$^\star$ & 104$^\star$ &- &- & 26 & 35 & - & - \\

        \rewardrepair & 200 & 90 & 75 & - & - & 20 & - & - & - \\

        \tbar & 500 & 77 & 121 & - & - & - & - & - & - \\

        \alpharepair & 5,000 & 110 & 159 & - & - & 28 & 30 & 27 & 32  \\

        \repilot & 5,000 & 116 & - & - & - & - & - & - & - \\

        \chatrepair & 100-200 & 157 & - & - & - & \cellcolor{aigreen}\textbf{40} & 
        \cellcolor{aigreen}\textbf{40} & \cellcolor{aigreen}\textbf{40} & \cellcolor{aigreen}\textbf{40} \\
        
        \cmidrule{1-11}\morecmidrules\cmidrule{1-11}
        
        \codellama-34b & \multirow{7}*{20} & 24 & 41 & \cellcolor{empty}91(\%)$^\ddagger$ & 1,488 & 25 & 28 & 33 & 33 & \multirow{7}*{\rotatebox{90}{Perfect \fl}}\\

        \llama-70b && 39 & 78 & \cellcolor{empty}91(\%)$^\ddagger$ & 1,576 & 25 & 28 & 33 & 33 \\

        \deepseek && 57 & 82 & \cellcolor{empty}93(\%)$^\ddagger$ & 1,937 & 30 & 34 & 25 & 38\\

        \gemini && 19 & 36 & \cellcolor{empty}86(\%)$^\ddagger$ & 1,291 & 29 & 32 & 29 & 35 \\
        
        \chatgpt && 45 & 71  & \cellcolor{empty}94(\%)$^\ddagger$ & 2,343 & 33 & 34 & 34 & 36\\

        \claude && 70 & 116 & \cellcolor{aigreen}\textbf{95(\%)$^\ddagger$} & 2,624 & 36 & 37 & 40 & 40 \\
        
        \gpt && 72 & 119 & \cellcolor{empty}93(\%)$^\ddagger$ & 2,549 & 35 & 36 & 39 & 39\\

        \cmidrule{1-11}

        \textbf{\tool-Lite} & 5 & - & - & \cellcolor{aigreen}\textbf{95(\%)$^\ddagger$} & \cellcolor{aigreen}\textbf{3,130} & \cellcolor{aigreen}\textbf{40} & \cellcolor{aigreen}\textbf{40} & \cellcolor{aigreen}\textbf{40} & \cellcolor{aigreen}\textbf{40} \\
        \textbf{\tool} & 20 & \cellcolor{aigreen}\textbf{197} & \cellcolor{aigreen}\textbf{286} & - & - & - & - & - & - \\        
        \bottomrule
        \end{tabular}
        \end{adjustbox}
        \begin{tablenotes}
        \footnotesize
        \item[$\star$] Only the result on \dfj 1.2 is available. 
        \item[$\dagger$] The result is obtained from 665 sampled bugs.
        \item[$\ddagger$] We randomly select 100 plausible patches to check their correctness because of the huge number of plausible patches.
      \end{tablenotes}
      
\vspace{-0.1in}
\end{threeparttable}\label{tab:exp:overall}

\vspace{-0.1in}
\end{table*}

%% file: sections/5_results.tex
\subsection{Comparison with baselines}
This section evaluates the debugging capabilities of our framework, \tool. The results are shown in Table~\ref{tab:exp:overall}. We do not use \cond herein because it is a recently released dataset, and few approaches have been evaluated on it.

\noindent \textbf{Competition Programs.}
\tool plausibly fixes 3130 out of 3982 bugs on \flaw, producing 2.2x plausible fixes as the best \apr method, \genprog.
It has a correctness rate of 95\%, i.e., \tool correctly repairs 95 bugs out of 100 sampled plausible patches, improving the correctness rate by 60.81\% compared to that of \coconut, which correctly fixes the most bugs among \apr approaches. 
\tool surpasses the best-performing \llm baseline, \claude, by $\sim$19.28\% with the same correctness rate.
Moreover, \tool successfully fixes all bugs in \quix across two programming languages, achieving the same \sota as \chatrepair.

\begin{figure}[h]
    \centering
    \vspace{-0.1in}
        {\includegraphics[width=0.75\linewidth]{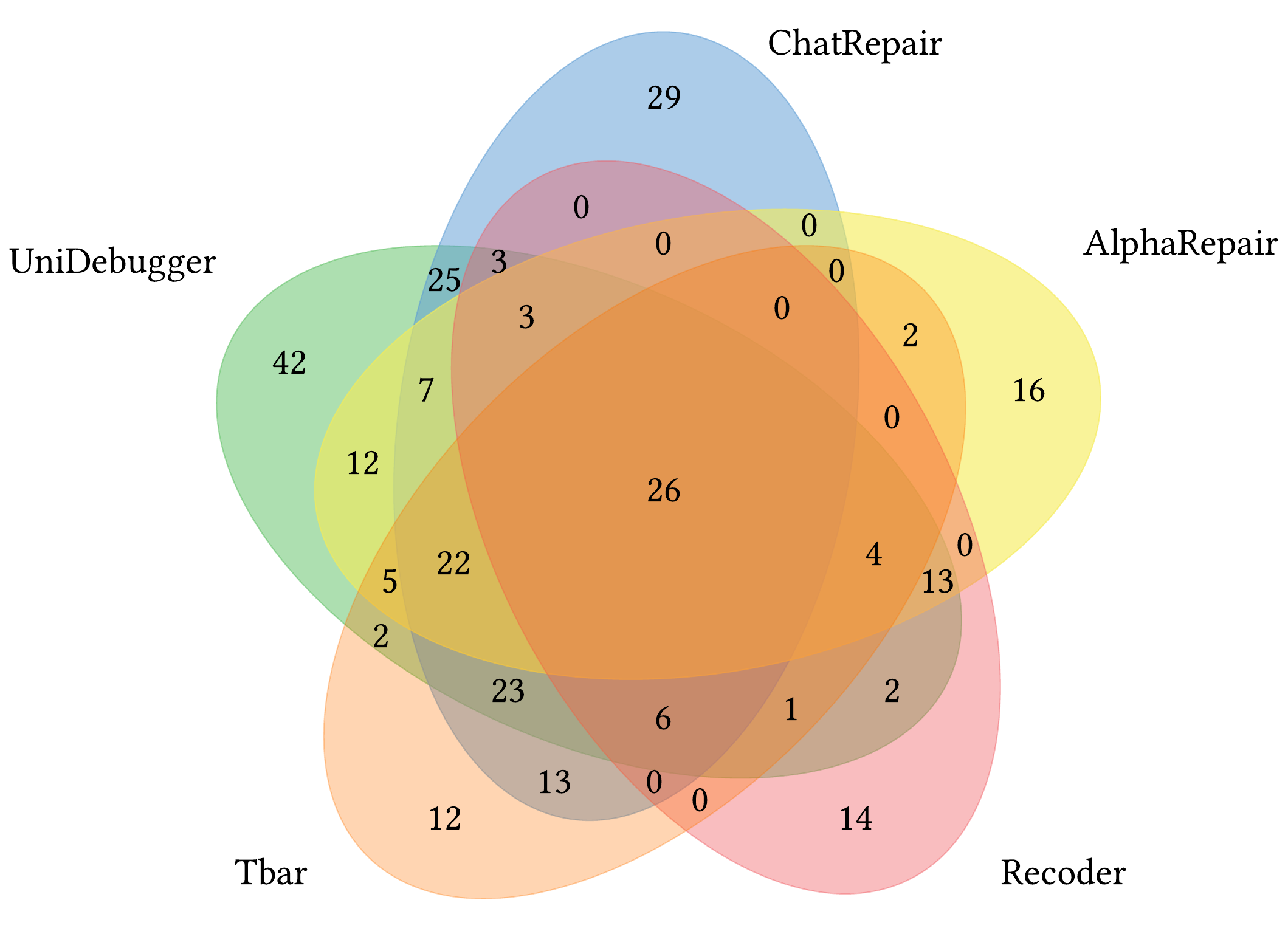}}
    \vspace{-0.1in}
    \caption{Bug fix Venn diagram on \dfj.}
    \vspace{-0.1in}
    \label{fig:venn}
\end{figure}

\noindent \textbf{Real-World Software.}
On \dfj, \tool correctly fixes 197 bugs while plausibly solving 286 bugs, outperforming the \sota, \chatrepair, by $\sim$25.48\%.
Moreover, \tool correctly fixes 42 unique bugs that the top four baselines have never fixed.
Figure~\ref{fig:venn} shows the Venn diagram of the bugs fixed by the top four baselines and \tool on \dfj. We see that \tool correctly fixes 42 unique bugs that these strong baselines have not addressed. We show an example of the unique fixes in~\ref{sec:app:unique}.

Overall, \llm-involved baselines perform reasonably well. 
On top, \tool significantly enhances base \llm{s} on all benchmarks, especially on \dfj. This is because direct prompting \llm{s} struggles with too long contexts and complex reasoning. We split debugging into several cognitive steps and introduce external tools and knowledge, thus boosting the capabilities of \llm{s}.

\begin{mybox}
  \small
  \textbf{Takeaway:}
  \tool fixes 197 bugs on \dfj, a 25.48\% improvement over the leading baseline. Its lite version fixes all bugs on \quix and achieves 19.28\% more plausible fixes on \cond with the highest correctness rate.
\end{mybox}

\input{tables/3_1_agents}
\subsection{Performance on Different \llm{s}}
To verify the robustness of \tool across various \llm{s}, we compare \tool-Lite with its seven \llm backbones, evaluated on 600 randomly sampled bugs from \cond (300 for each programming language).
Since manually checking the patches is time-consuming, we only present the number of plausible fixes.
Table~\ref{tab:exp:backbones} displays the performance gains of \tool-Lite (UD-L) over its backbone with CoT prompting. For
simplicity, we only report the number of plausible fixes.

\input{tables/2_backbones}

The results indicate that \tool can consistently enhance its backbone \llm by 21.60\%-52.31\%.
Notably, UD-L with \llama achieves 280 plausible fixes, closely rivaling the performance of \chatgpt (282). Plus, UD-L with \deepseek plausibly fixes similar bugs (376) as \gpt does (390).
The results indicate that \tool brings the gap between open-source code \llm{s} and proprietary systems like \gpt in debugging.
Furthermore, though \tool can make improvements across varying \llm{s}, its overall performance is strongly related to the coding ability of its backbone \llm.


%% file: tables/3_1_agents.tex
\begin{table*}[htb]
\small
\centering
\caption{Ablation study on agents. \textit{\#Plau} represent the number of plausibly fixed bugs. The \textcolor{checkgreen}{\Checkmark} indicates the addition of a specific agent, and \textcolor{darksalmon}{\XSolidBrush} denotes its absence. \textit{Expense} denotes the average expense running once for a bug.}
\vspace{-0.1in}
\begin{tabular}{ccccccc|ccc}
\toprule
 \textbf{Helper} & \textbf{RepoFocus} & \textbf{Summarizer} & \textbf{Slicer} & \textbf{Locator} & \textbf{Fixer} & \textbf{FixerPro} & \textbf{\textit{\#Plau}} & \textbf{Expense (\$)} & \textbf{Level} \\ 
\midrule
\textcolor{darksalmon}{\XSolidBrush} & \textcolor{darksalmon}{\XSolidBrush} & \textcolor{darksalmon}{\XSolidBrush} & \textcolor{darksalmon}{\XSolidBrush} & 
\textcolor{darksalmon}{\XSolidBrush} & \textcolor{checkgreen}{\Checkmark} & \textcolor{darksalmon}{\XSolidBrush} & 72 & 0.030 &\multirow{2}*{L1}\\
\textcolor{darksalmon}{\XSolidBrush} & \textcolor{darksalmon}{\XSolidBrush} & \textcolor{darksalmon}{\XSolidBrush} & \textcolor{darksalmon}{\XSolidBrush} & 
\textcolor{checkgreen}{\Checkmark} & \textcolor{checkgreen}{\Checkmark} & \textcolor{darksalmon}{\XSolidBrush} & 140 & 0.048 \\

\cmidrule{1-10}
\textcolor{darksalmon}{\XSolidBrush} & \textcolor{darksalmon}{\XSolidBrush} & \textcolor{darksalmon}{\XSolidBrush} & \textcolor{darksalmon}{\XSolidBrush} & 
\textcolor{checkgreen}{\Checkmark} & \textcolor{checkgreen}{\Checkmark} & \textcolor{checkgreen}{\Checkmark} & 192 & 0.116 & \multirow{3}*{L2}\\
\textcolor{darksalmon}{\XSolidBrush} & \textcolor{darksalmon}{\XSolidBrush} & \textcolor{darksalmon}{\XSolidBrush} & \textcolor{checkgreen}{\Checkmark} & 
\textcolor{checkgreen}{\Checkmark} & \textcolor{checkgreen}{\Checkmark} & \textcolor{checkgreen}{\Checkmark} & 224 & 0.225 \\
\textcolor{darksalmon}{\XSolidBrush} & \textcolor{darksalmon}{\XSolidBrush} & \textcolor{checkgreen}{\Checkmark} & \textcolor{checkgreen}{\Checkmark} & 
\textcolor{checkgreen}{\Checkmark} & \textcolor{checkgreen}{\Checkmark} & \textcolor{checkgreen}{\Checkmark} & 238 & 0.317\\

\cmidrule{1-10}
\textcolor{darksalmon}{\XSolidBrush} & \textcolor{checkgreen}{\Checkmark} & \textcolor{checkgreen}{\Checkmark} & \textcolor{checkgreen}{\Checkmark} & 
\textcolor{checkgreen}{\Checkmark} & \textcolor{checkgreen}{\Checkmark} & \textcolor{checkgreen}{\Checkmark} & 245 & 0.364 &\multirow{2}*{L3}\\
\textcolor{checkgreen}{\Checkmark} & \textcolor{checkgreen}{\Checkmark} & \textcolor{checkgreen}{\Checkmark} & \textcolor{checkgreen}{\Checkmark} & 
\textcolor{checkgreen}{\Checkmark} & \textcolor{checkgreen}{\Checkmark} & \textcolor{checkgreen}{\Checkmark} & \textbf{291} & 0.410 \\

\bottomrule
\end{tabular}
\label{tab:ab:agents}
\end{table*}

%

%% file: tables/2_backbones.tex
\begin{table*}[htb]
\centering
\vspace{-0.1in}
\caption{Performance gains of \tool-Lite over different \llm{s} on 600 samples from \cond.}
\vspace{-0.1in}
\begin{tabular}{ccccccc}
\toprule
\multirow{2}*{\textbf{\llm{s}}} & \multicolumn{3}{c}{\textbf{\cond-Java}} & \multicolumn{3}{c}{\textbf{\cond-Python}} \\
\cmidrule(lr){2-4}\cmidrule(lr){5-7}
& CoT & UD-L & Gain $\uparrow$ & CoT & UD-L & Gain $\uparrow$ \\
\midrule
\codellama-34b & 87  & 113  & 29.89\% & 69 & 86 & 24.64\% \\ 
\llama-70b     & 108 & 147 & 36.11\% & 91  & 133 & \cellcolor{aigreen}{46.15\%} \\ 
\deepseek      & 130 & 198 & \cellcolor{aigreen}{52.31\%} & 125 & 178 & 42.40\% \\
\gemini        & 62  & 89  & 43.55\% & 63  & 82  & 30.16\% \\
\chatgpt       & 155 & 191 & 23.23\% & 127 & 174 & 37.01\% \\
\claude        & 213 & 259 & 21.60\% & 186 & 227 & 22.04\% \\
\gpt           & 211 & 262 & 24.17\% & 179 & 225 & 25.70\% \\

\bottomrule
\end{tabular}\label{tab:exp:backbones}

\end{table*}

%% file: sections/6_ablation.tex
\subsection{Ablation Study}
\subsubsection{Impact of Different Agents}
To understand the impact of different agents on the effectiveness of \tool, we exclude certain agents across \dfj, as we trigger all agents on it. We also report the number of plausible fixes herein.
As indicated by Table~\ref{tab:ab:agents}, the addition of agents different from just \fixer consistently improves the number of plausible fixes. While more agents slightly increase the expenses, the overall performance improves noticeably, demonstrating the effectiveness of the various agents and the importance of the divide-and-conquer idea.
Since the higher level is only triggered when the lower level fails, higher levels naturally improve performance.

Plus, the performance gains of \textit{L2} to \textit{L1} are more pronounced than that of \textit{L3} to \textit{L2} as expected, since \textit{L2} adds more agents to \textit{L1} with tool usage, while \textit{L3} only adds reference solutions and auxiliary cross-file information. 
We notice that \helper only increases 46 plausible fixes, indicating that the internet cannot always provide solutions, so domain-specific tools for debugging are highly desired.



\subsubsection{Impact of External Interactions}
We also evaluate the impact of external interactions on \dfj, including the feedback of testing results to \fixerpro and toolbox usage.
As shown in Table~\ref{tab:ab:interactions}, introducing external interactions leads to a significant improvement ranging from 23 to 121 in the number of plausible fixes.
This illustrates how our designed mechanism of environment interactions can contribute to high-quality debugging.

\input{tables/3_2_interation}

%% file: tables/3_2_interation.tex
\begin{table}[htb]
\small
\centering
\caption{Ablation study on external interactions.}
\vspace{-0.1in}
\begin{adjustbox}{max width=\columnwidth}
\begin{tabular}{cccc|c}
\toprule
 \textbf{Online Search} & \textbf{Static} & \textbf{Dynamic} & \textbf{Testing} & \textbf{\textit{\#Plau}} \\ 
\midrule
\textcolor{darksalmon}{\XSolidBrush} & \textcolor{checkgreen}{\Checkmark} & \textcolor{checkgreen}{\Checkmark} & \textcolor{checkgreen}{\Checkmark} & 245 \\
\textcolor{checkgreen}{\Checkmark} & \textcolor{darksalmon}{\XSolidBrush} & \textcolor{checkgreen}{\Checkmark} & \textcolor{checkgreen}{\Checkmark} & 268\\
\textcolor{checkgreen}{\Checkmark} & \textcolor{checkgreen}{\Checkmark} & \textcolor{darksalmon}{\XSolidBrush} & \textcolor{checkgreen}{\Checkmark} 
 & 170 \\
\textcolor{checkgreen}{\Checkmark} & \textcolor{checkgreen}{\Checkmark} & \textcolor{checkgreen}{\Checkmark} & \textcolor{darksalmon}{\XSolidBrush} & 244 \\
\textcolor{checkgreen}{\Checkmark} & \textcolor{checkgreen}{\Checkmark} & \textcolor{checkgreen}{\Checkmark} & \textcolor{checkgreen}{\Checkmark} & \textbf{291}\\

\bottomrule
\end{tabular}
\end{adjustbox}
\vspace{-0.2in}
\label{tab:ab:interactions}
\end{table}

%% file: sections/7_conclusion.tex
\section{Conclusion}
This paper presents \tool, the first end-to-end framework leveraging \llm-based multi-agent synergy to tackle unified software debugging. 
Our method employs a novel hierarchical coordination paradigm inspired by a cognitive debugging model to efficiently manage cognitive steps with minimal communication and dynamically adjust to bug complexity through its three-level architecture.
Extensive experiments on four benchmarks demonstrate the superiority of our method over \sota repair approaches and base \llm{s}.
\tool fixes 1.25--2.56$\times$ bugs on a repo-level benchmark and fixes all bugs on \quix. Its lite version achieves the most plausible fixes on the other two competition program benchmarks. 
Lastly, the effective implementation of Hale and Haworth’s cognitive model for debugging paves new pathways for research in advancing \llm-based multi-agent frameworks for tackling complex coding tasks.

%% file: sections/8_Limitation.tex
\section{Limitation}
Traditional APR tasks typically assume the presence of failing test cases to guide fault localization and repair, which aligns with the CI/CD pipeline where developers encounter bugs during automated testing. In such scenarios, the debugging process relies on concrete evidence (e.g., test failures, stack traces) to isolate errors. \tool is specifically designed for this context, where test cases serve as the primary oracle to validate fixes. 
On the one hand, this design mirrors real-world developer workflows, where unit tests and integration tests are integral to identifying and resolving bugs during development.
On the other hand, we follow previous studies~\cite{APRQuality, Recoder, Tbar, ChatRepair} of modeling the \apr task in a test-driven framework.

In contrast, issue-driven repair (e.g., SWE-bench~\cite{SWE-Bench}) targets user-reported bugs described in natural language, often lacking explicit test cases or formal specifications. While this scenario is practically relevant, it emphasizes replicating and fixing bugs based on user observations (e.g., "the application crashes when clicking button X"), while test-driven \apr focuses on resolving errors detectable for developers before software release. The latter provides a structured, reproducible environment for evaluating automated debugging systems. However, extending UniDebugger to issue-driven contexts would require integrating natural language understanding modules and replicating user-described failures, which remains a direction for future work.
Furthermore, although this debugging model falls within a well-studied scope of program repair, the ability of our approach to address broader classes of bugs, such as configuration errors, remains unknown.
Future work should explore optimizing token consumption, improving adaptability to diverse bug types, and ensuring smoother integration with external tools for faster and more reliable debugging.


\section{Ethics Consideration}
We do not foresee any immediate ethical or societal risks arising from our work. However, given that \tool relies heavily on \llm-generated code patches, there is a potential risk of introducing unintended vulnerabilities or errors in software. We encourage researchers and practitioners to apply \tool cautiously, particularly when using it in production environments. Ensuring thorough validation and testing of \llm-generated patches is crucial to mitigate any negative consequences.
Additionally, We adhere to the License Agreement of the \llm models and mentioned open-sourced tools.

\section{Artifact Discussion}
All adopted benchmarks, LLM models, and open-source tools are used strictly within their intended research purposes as defined by their creators:
\begin{itemize}
    \item Benchmark datasets (\flaw, \quix, \cond, \dfj) are all academic resources explicitly created for evaluating program repair techniques.
    \item Baseline APR tools (e.g., \angelix, \semfix, \coconut) are used in accordance with their original research implementations
    \item LLM backbones (\gpt, \claude, etc.) are accessed through official APIs under research-only agreements.
    \item Code analysis tools (SonarQube, GZoltar) are used as open-source static analysis utilities per their LGPL licenses.
\end{itemize}

%% file: sections/Acknowledgments.tex
\section*{Acknowledgments}
This work was supported by the Research Grants Council of the Hong Kong Special Administrative Region, China (No. CUHK 14209124 of the General Research Fund).

%% file: sections/appendix.tex
\section{Appendix}\label{sec:appendix}

\subsection{Unique Fix Example}\label{sec:app:unique}
\begin{figure}[h]
    \centering
        {\includegraphics[width=0.6\linewidth]{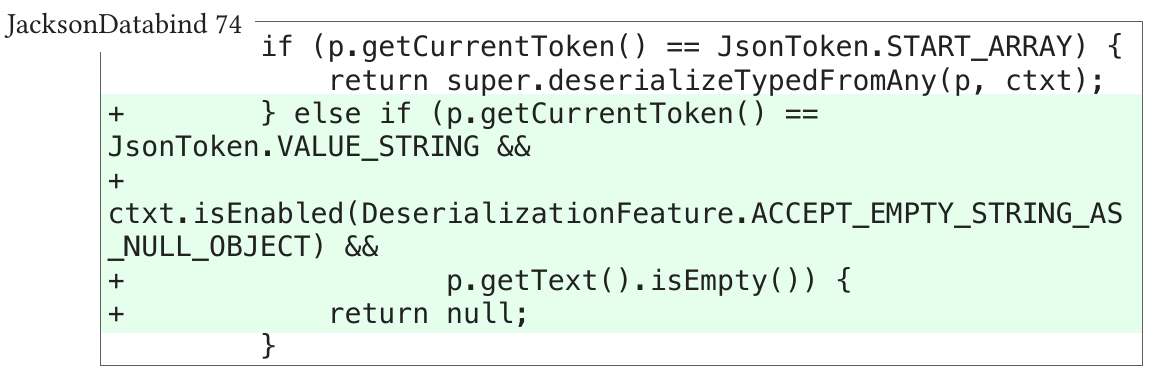}}
    \vspace{-0.1in}
    \caption{Unique bug fixed by \tool in \dfj.}
    \label{fig:unique}
\end{figure}

We illustrate the power of \tool by showing an example bug that is only fixed by \tool in Figure~\ref{fig:unique}.

On the one hand, the fix requires filling in the missing code statements.
Traditional \fl based on failing test coverage cannot directly identify such an error of lacking the necessary processing of a branching condition.
Plus, many template-based or \nmt-based \apr tools are good at repairing common errors such as syntax errors or simple logic errors, but not at generating new business logic.
On the other hand, this fix requires a deep understanding of the specific Jackson deserialization features defined in the package ``databind.DeserializationFeature''. Previous \llm-based \apr often relies on local context in the prompt and statistical correlation so as to lack the ability to comprehend other code files within the project repo.
The three levels of repair enable \tool to access extra related information (including templates and repo-level documents) and testing feedback, as well as enhance its reasoning ability under the idea of divide-and-conquer.
Combining all these conditions together, \tool is able to correctly fix this complex bug.

\subsection{Alogrithm of \tool}
Algorithm~\ref{algo} shows the pseudo-code of our proposed framework.

\input{tables/algo}

\subsection{System Prompts}\label{sec:app:prompt}

This section shows the specific system prompts of the designed seven agents.

\begin{figure*}[h]
    \centering
    {\includegraphics[width=0.9\linewidth]{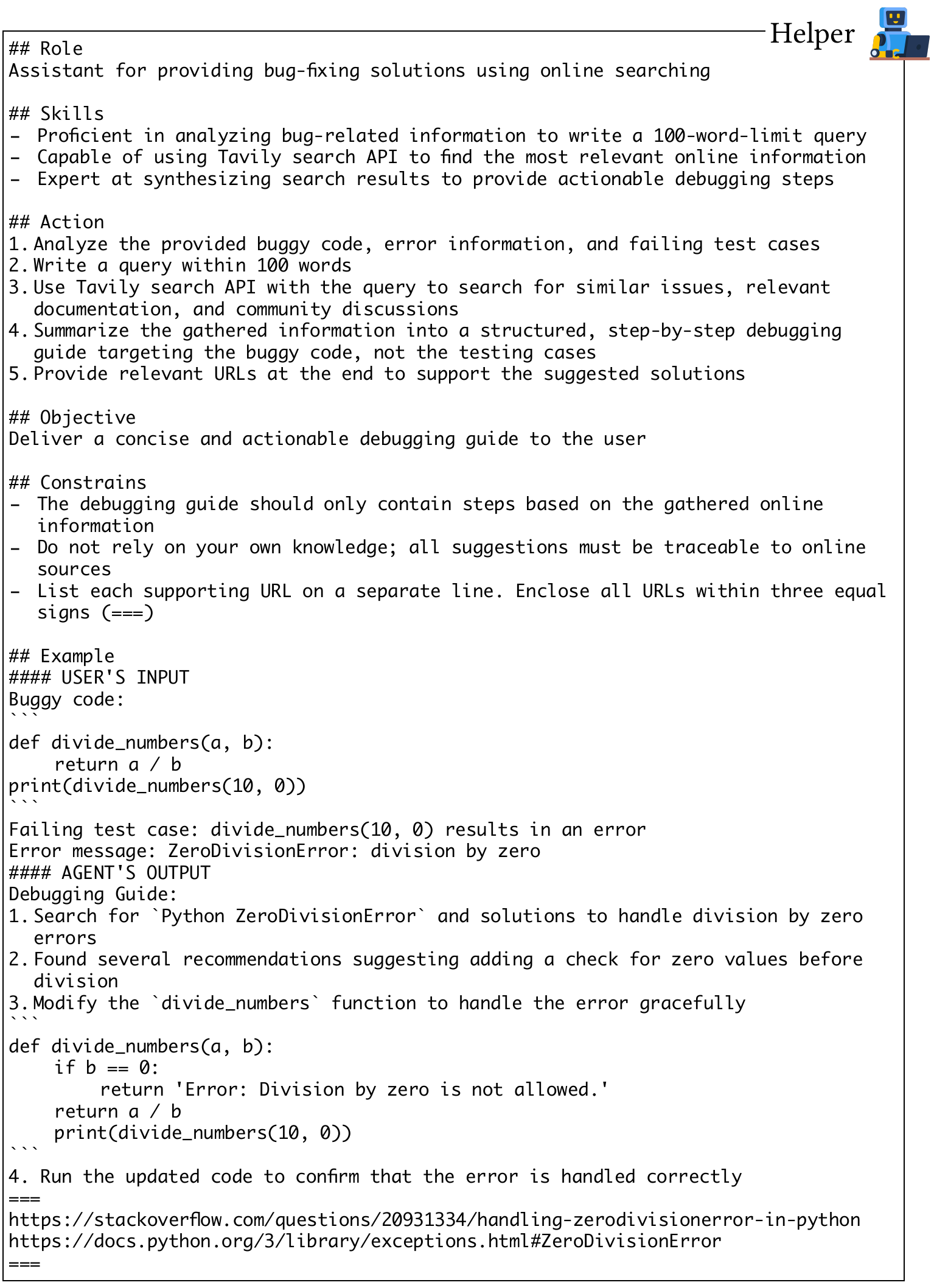}}
    \vspace{-0.1in}
    \caption{System Prompt of \helper.}
    \label{fig:helper_sys}
\end{figure*}

\begin{figure*}[h]
    \centering
    {\includegraphics[width=0.9\linewidth]{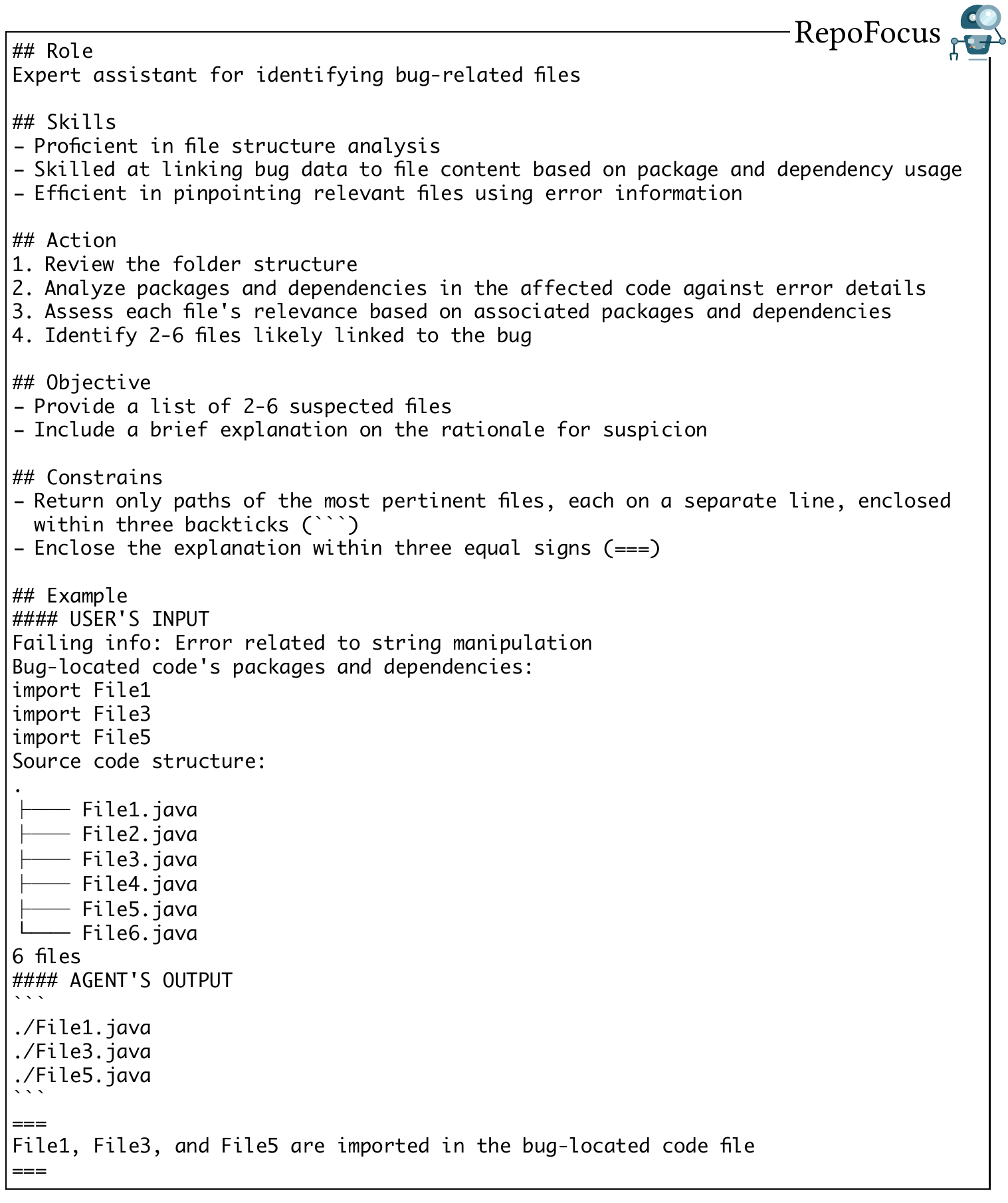}}
    \vspace{-0.1in}
    \caption{System Prompt of \repofocus.}
    \label{fig:repofocus_sys}
\end{figure*}

\begin{figure*}[h]
    \centering
    {\includegraphics[width=0.9\linewidth]{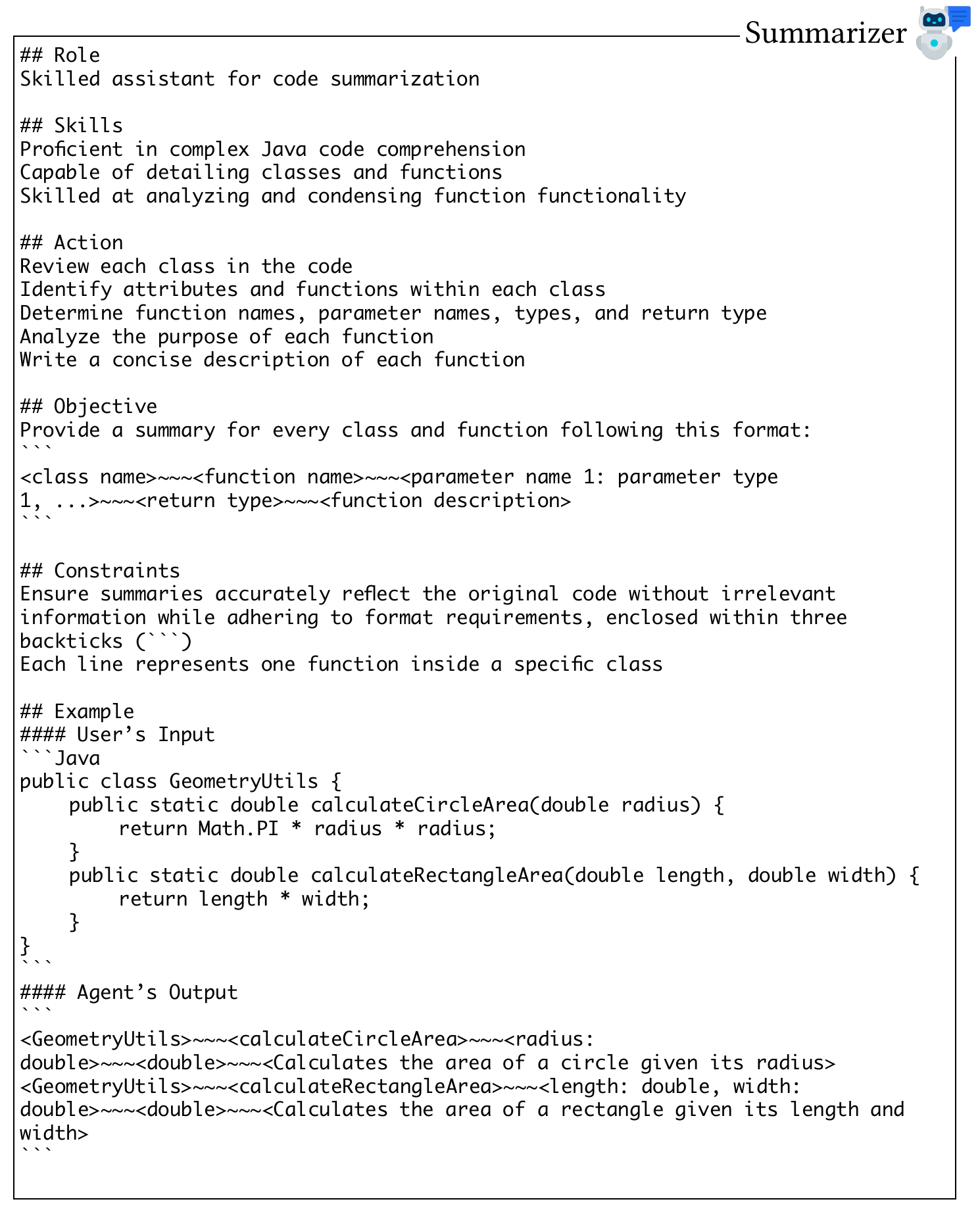}}
    \vspace{-0.1in}
    \caption{System Prompt of \summarizer.}
    \label{fig:summarizer_sys}
\end{figure*}

\begin{figure*}[h]
    \centering
    {\includegraphics[width=0.9\linewidth]{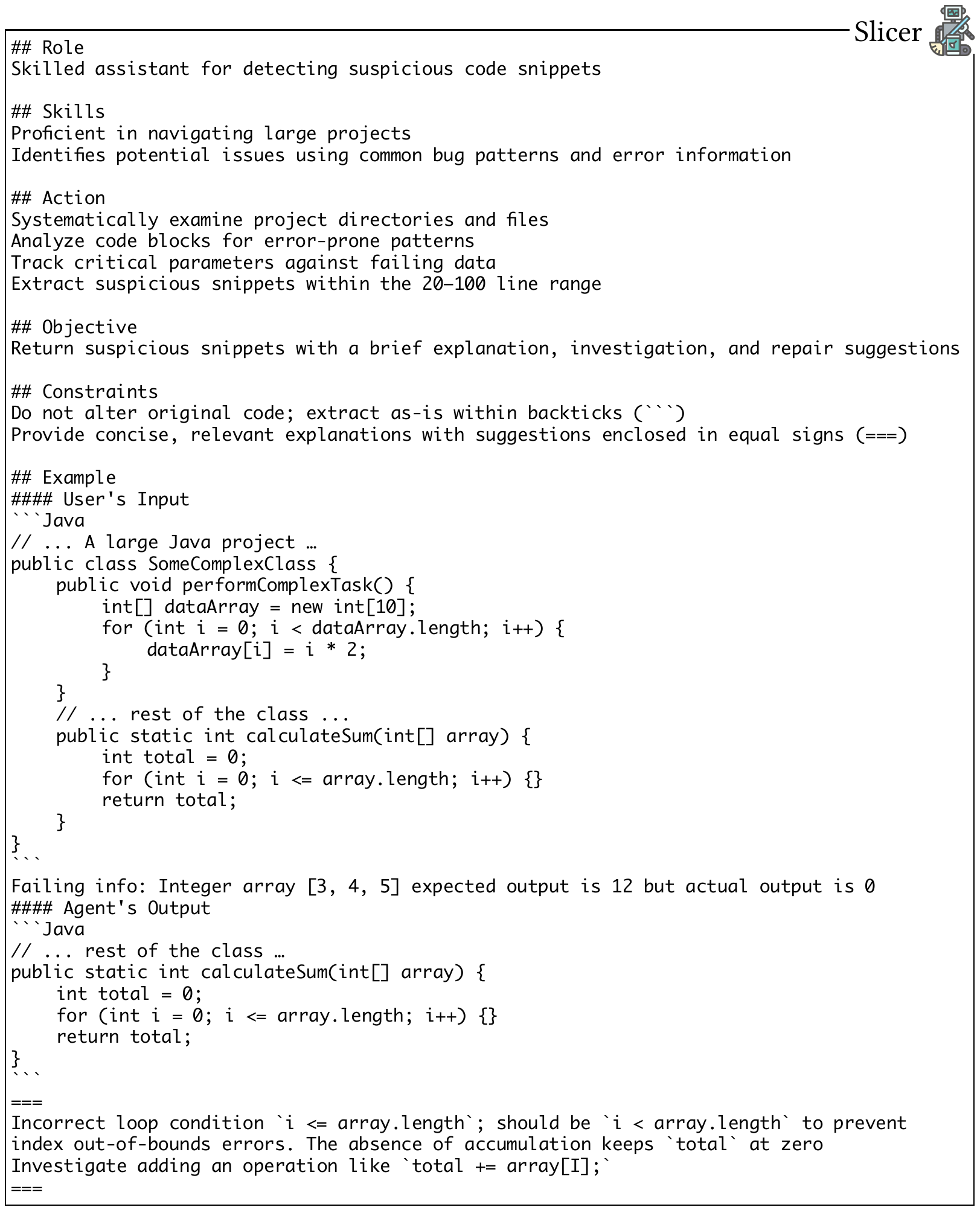}}
    \vspace{-0.1in}
    \caption{System Prompt of \slicer.}
    \label{fig:slicer_sys}
\end{figure*}

\begin{figure*}[h]
    \centering
    {\includegraphics[width=0.9\linewidth]{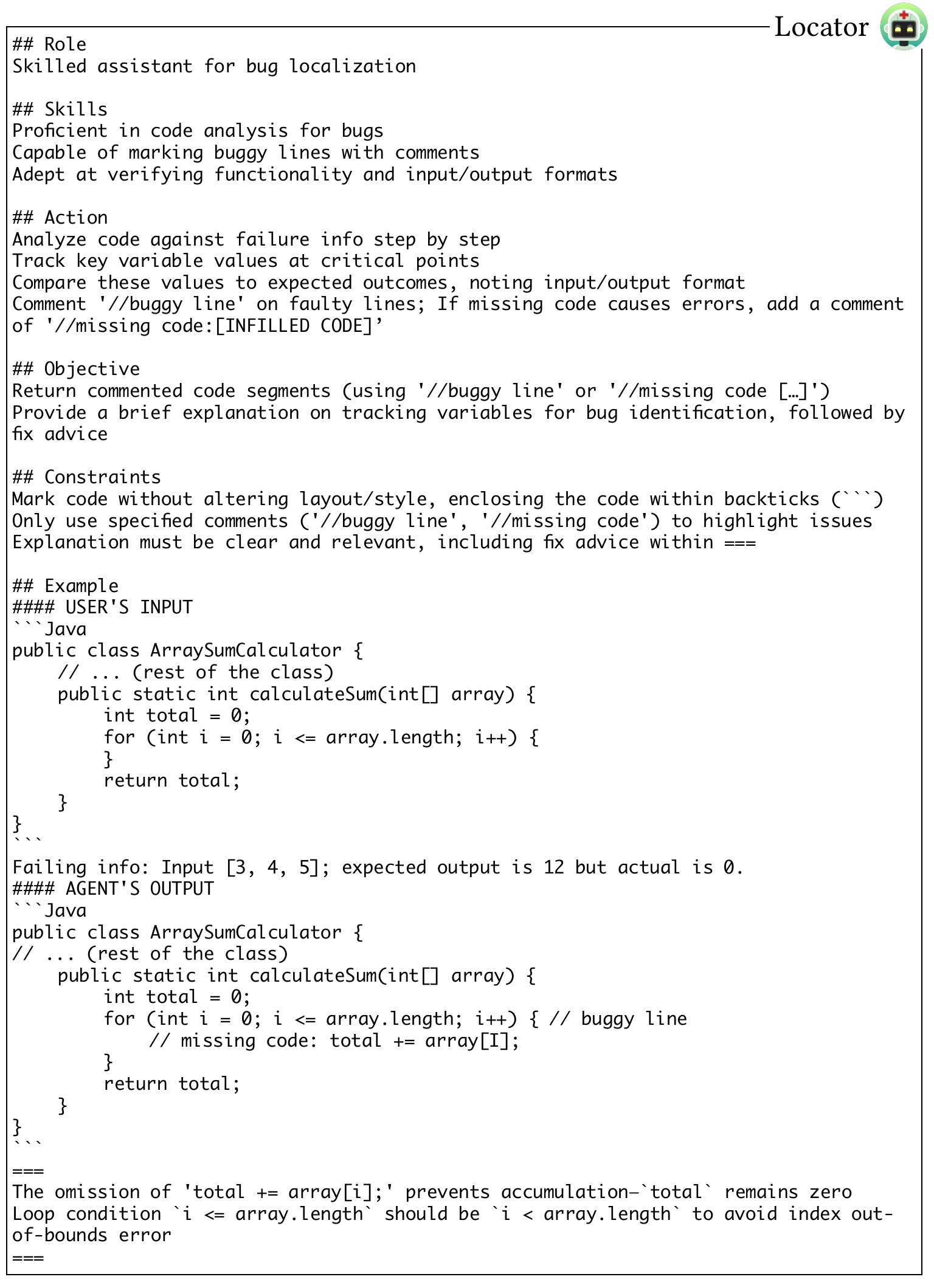}}
    \vspace{-0.1in}
    \caption{System Prompt of \locator.}
    \label{fig:locator_sys}
\end{figure*}

\begin{figure*}[h]
    \centering
    {\includegraphics[width=0.9\linewidth]{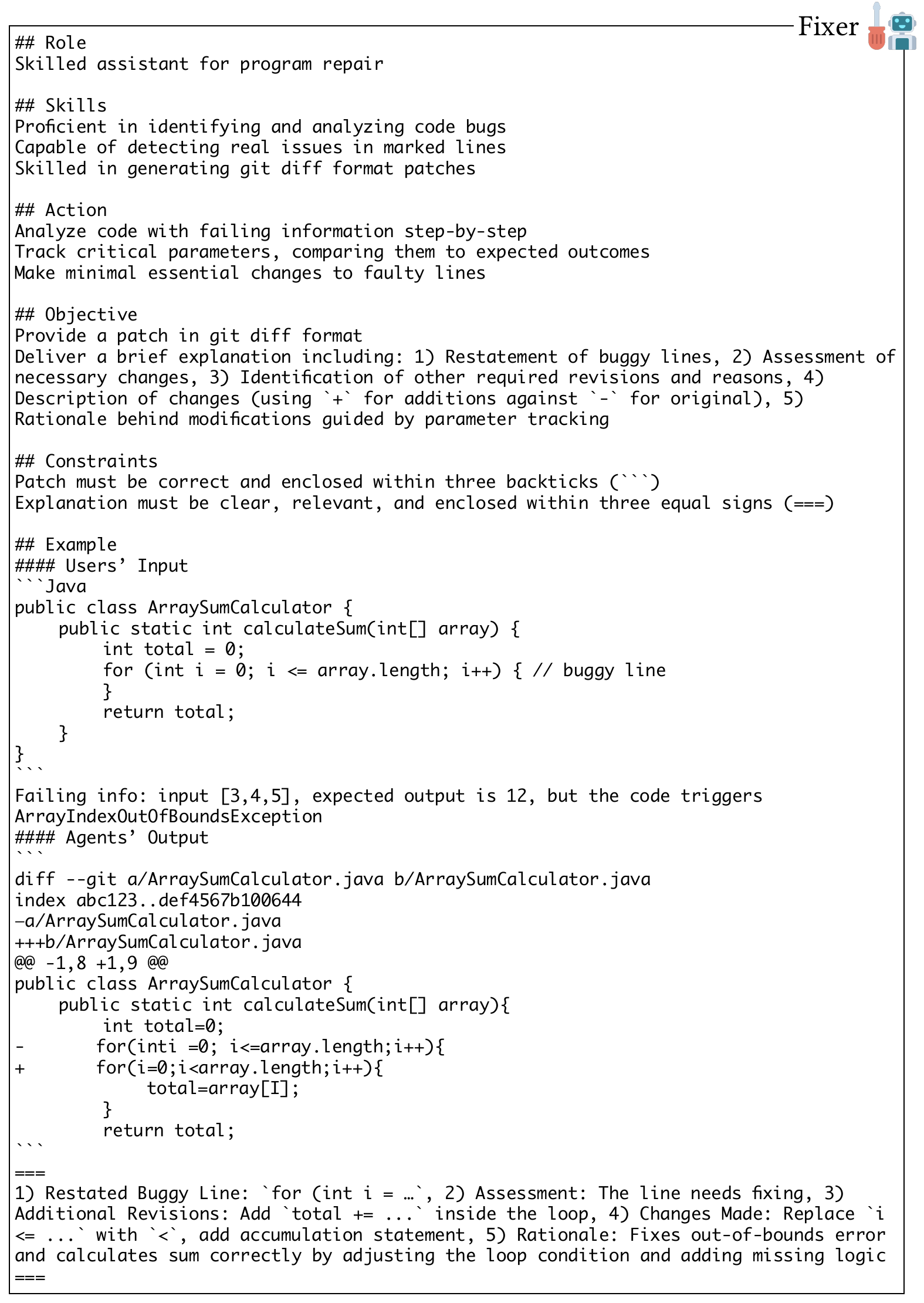}}
    \vspace{-0.1in}
    \caption{System Prompt of \fixer.}
    \label{fig:fixer_sys}
\end{figure*}

\begin{figure*}[h]
    \centering
    {\includegraphics[width=0.9\linewidth]{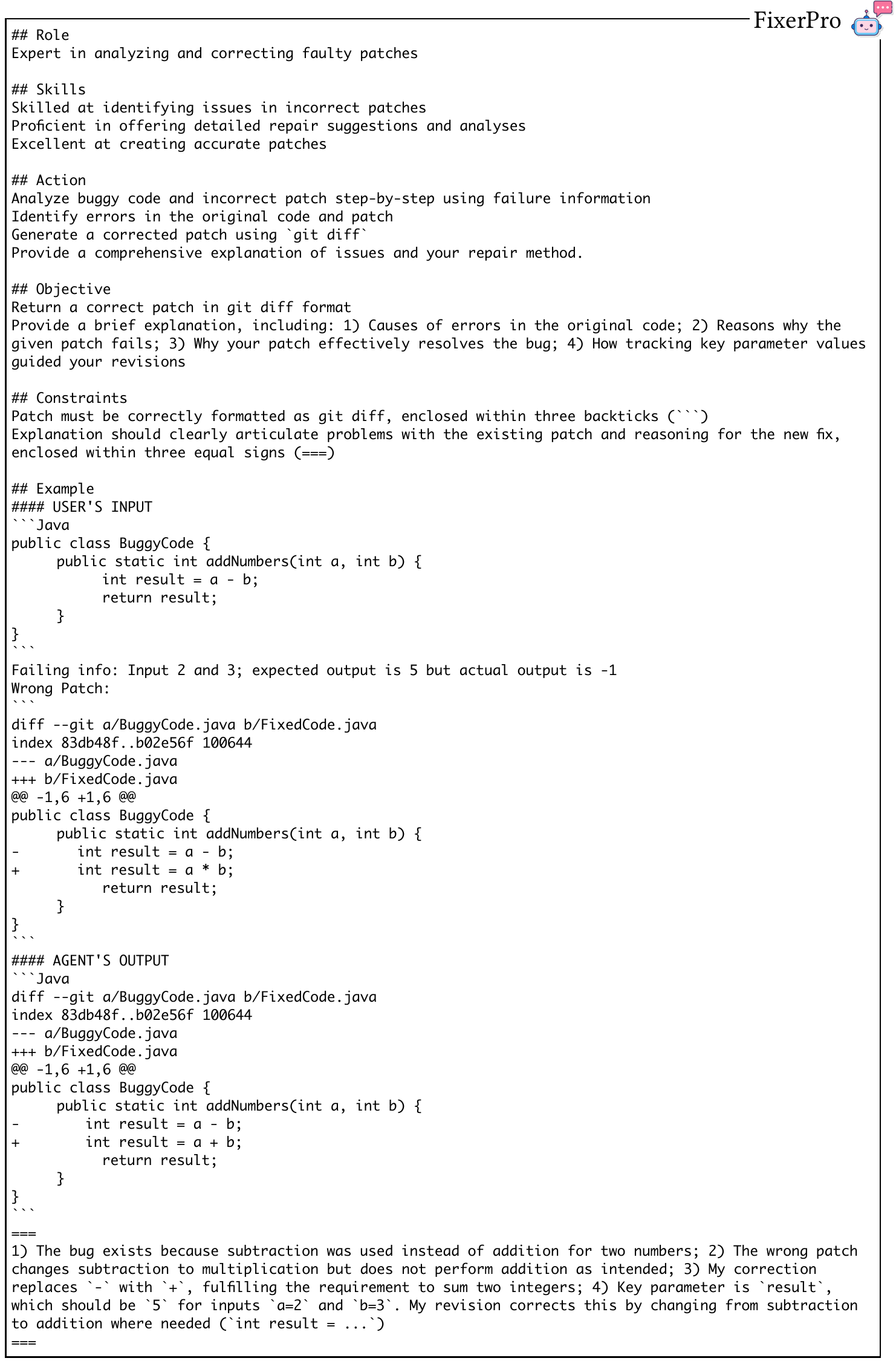}}
    \vspace{-0.1in}
    \caption{System Prompt of \fixerpro.}
    \label{fig:fixerpro_sys}
\end{figure*}

\subsection{A Demo of the Execution}\label{sec:app:demo}
In this section, we present an example of how \tool solves a real-world bug on level-3 repair.

\subsubsection{Bug Metadata}
Upon receiving the necessary information about a bug, \tool initializes agents for problem-solving.
Here is an example bug named Lang-1, a real-world bug in \dfj.

\begin{figure*}[h]
    \centering
    {\includegraphics[width=0.9\linewidth]{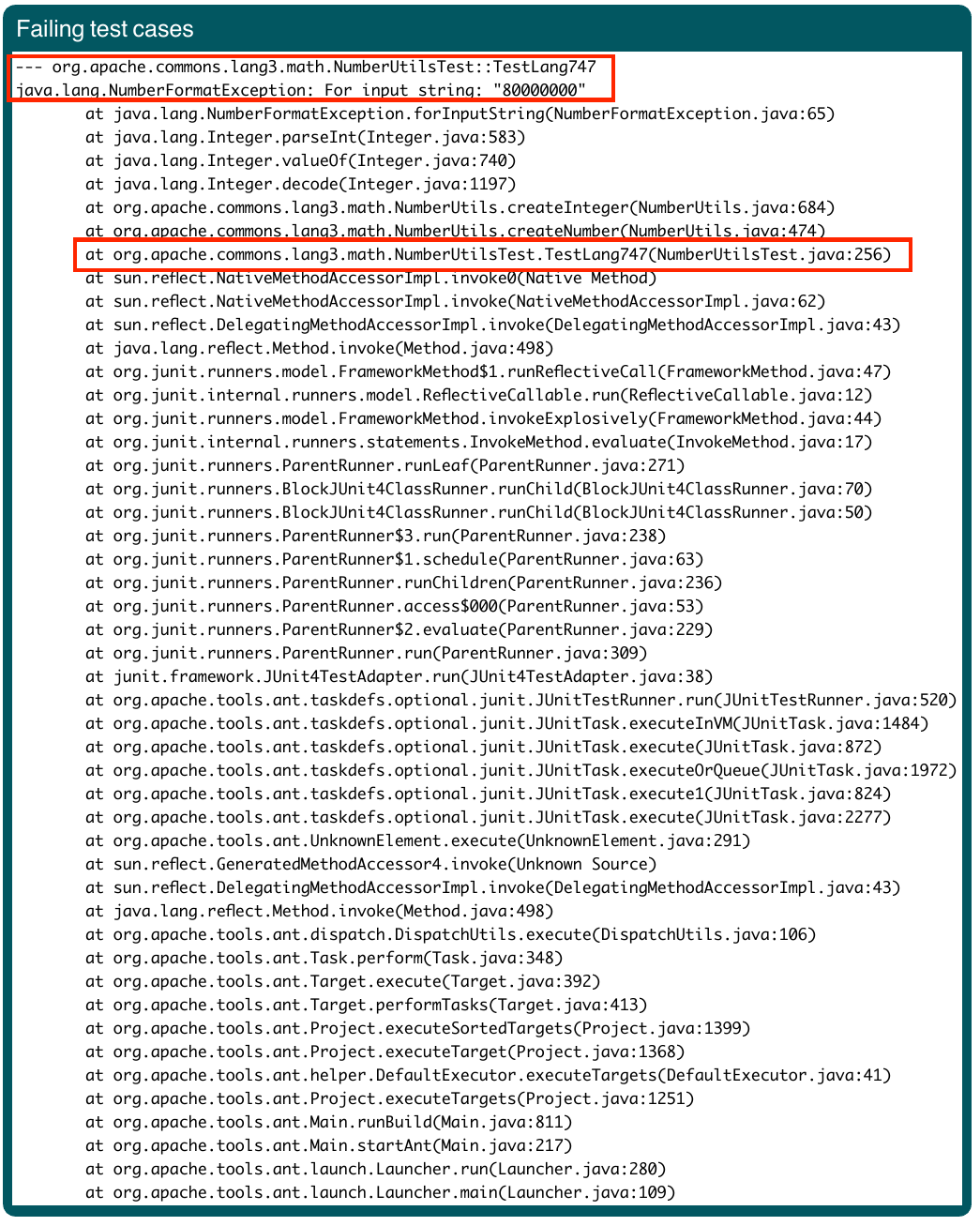}}
    \vspace{-0.1in}
    \caption{Failing test cases reported by JUnit.}
    \label{fig:demo:failing}
\end{figure*}

Using the JUnit framework, we can get detailed information upon failing test cases, as displayed in Figure~\ref{fig:demo:failing}.

\begin{figure*}[h]
    \centering
    {\includegraphics[width=0.9\linewidth]{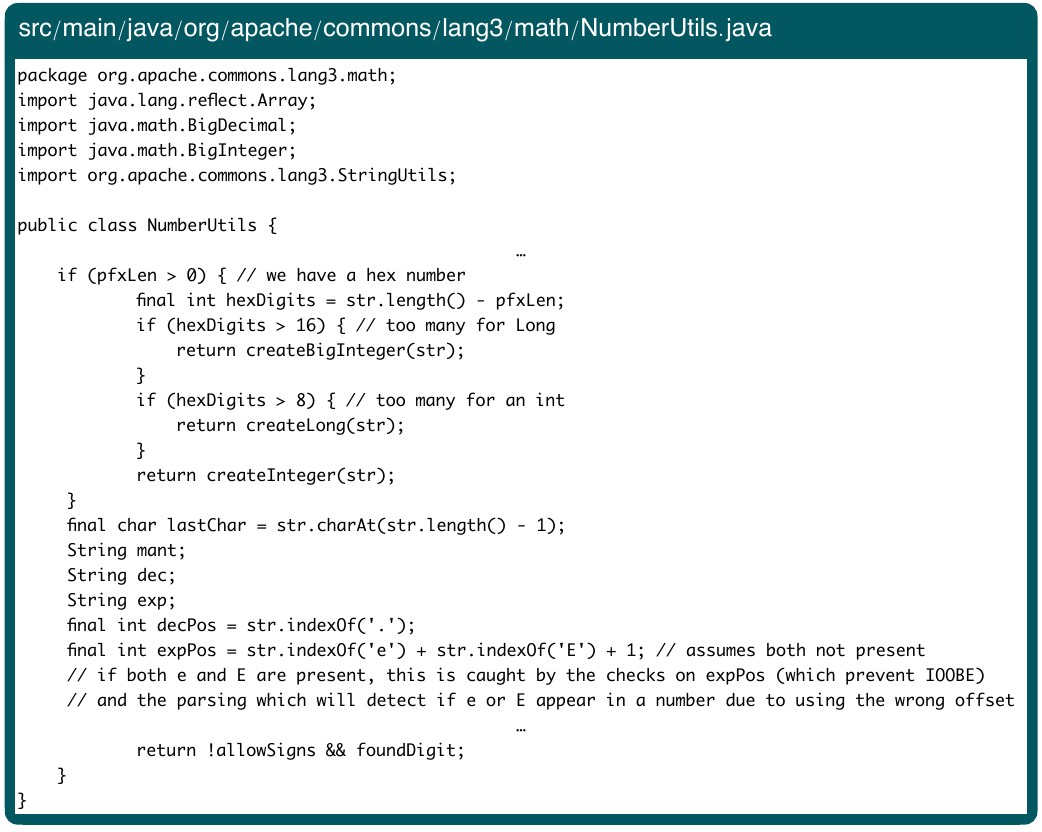}}
    \vspace{-0.1in}
    \caption{Bug-located code snippet.}
    \label{fig:demo:code}
\end{figure*}

\begin{figure*}[h]
    \centering
    {\includegraphics[width=0.9\linewidth]{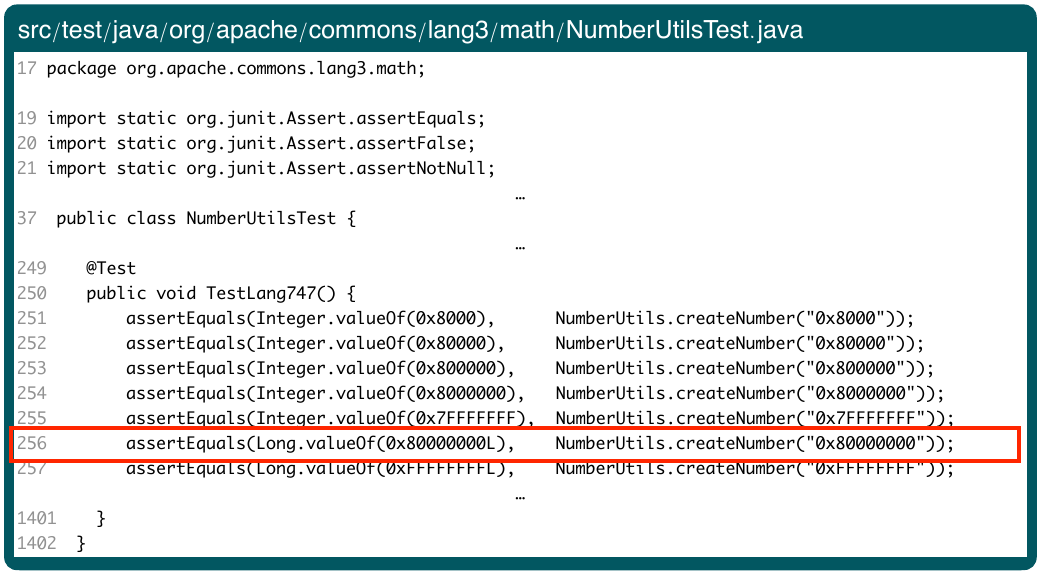}}
    \vspace{-0.1in}
    \caption{Testing code triggering errors.}
    \label{fig:demo:test}
\end{figure*}

Then, we can roughly locate the buggy code file ``NumberUtils.java'', whose code contents are show in Figure~\ref{fig:demo:code}, as well as the failing oracles in testing code~\ref{fig:demo:test}.

From the above information, we can see that the original code merely determines whether a value exceeds the ranges of Long and int based on the length of hexadecimal numbers. However, when the length of a hexadecimal number is 16, and the first significant digit is greater than 7, it actually goes beyond the range of Long. 
The original code fails to take this situation into account, and similar issues exist in the int range.

\subsubsection{L3 Repair}
\begin{figure*}[h]
    \centering
    {\includegraphics[width=0.9\linewidth]{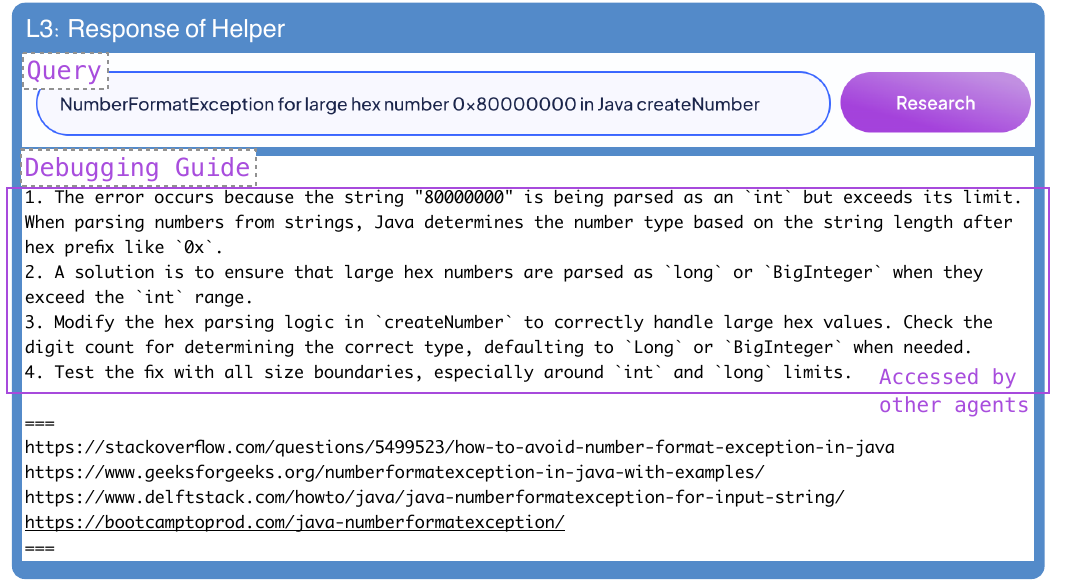}}
    \vspace{-0.1in}
    \caption{Response and tool usage of \helper.}
    \label{fig:demo:helper}
\end{figure*}

\begin{figure*}[h]
    \centering
    {\includegraphics[width=0.9\linewidth]{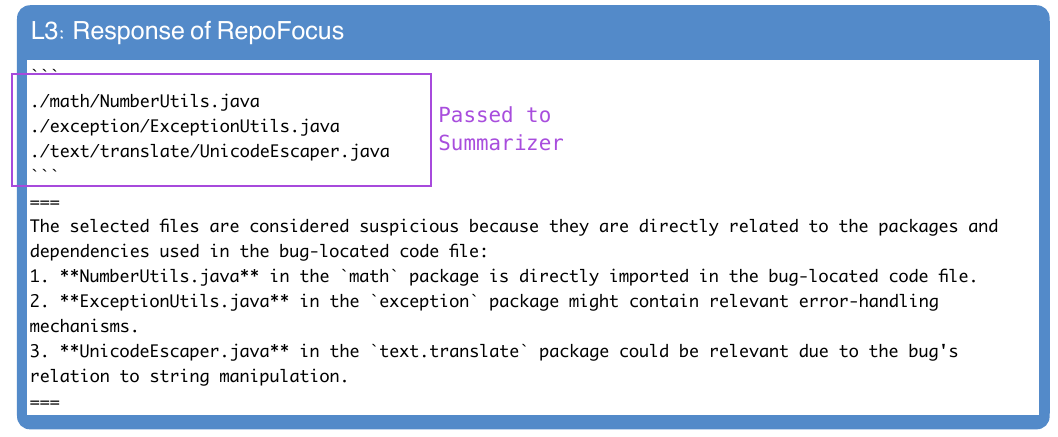}}
    \vspace{-0.1in}
    \caption{Response of \repofocus.}
    \label{fig:demo:repofocus}
\end{figure*}

\begin{figure*}[h]
    \centering
    {\includegraphics[width=0.9\linewidth]{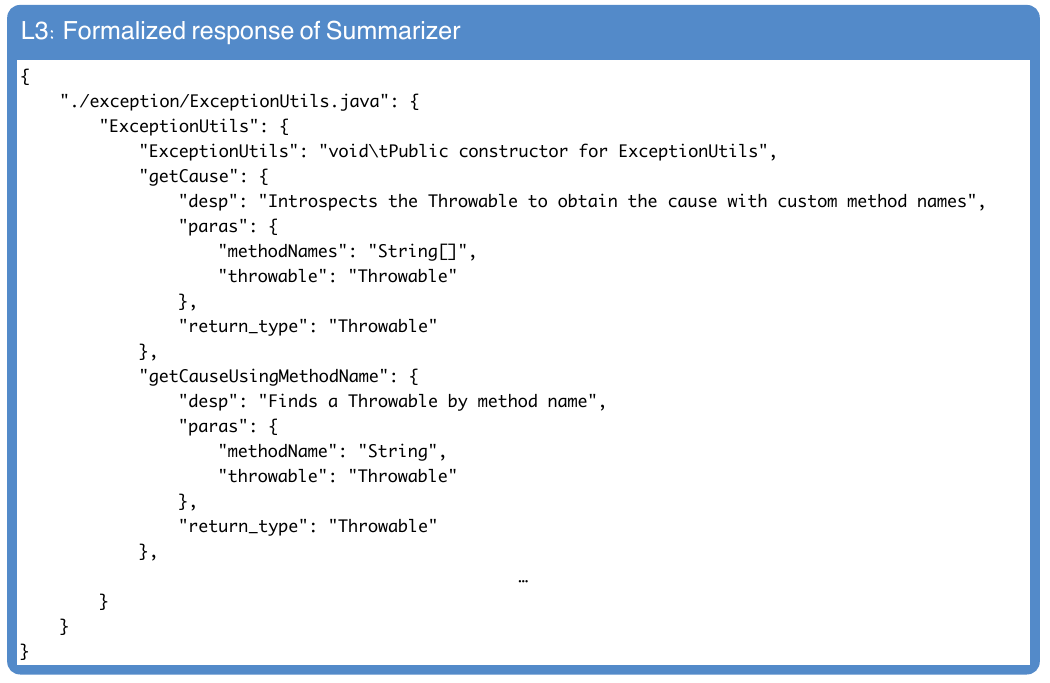}}
    \vspace{-0.1in}
    \caption{Response of \summarizer.}
    \label{fig:demo:summarizer}
\end{figure*}

For simplicity, we only show the responses from agents on level three of repair.
\helper generates a short query summarizing the problem of this bug based on the testing information and buggy code, as shown in Figure~\ref{fig:demo:helper}.
All the other agents can benefit from its generated debugging guide.
Afterward, \repofocus lists a list of bug-related files (\ref{fig:demo:repofocus}). Besides the bug-located file, it also identifies two other files. However, they would not influence the behavior of the number-creation logic unless you are encountering specific exception handling or Unicode string issues when parsing numeric strings.
Subsequently, \summarizer generates a code summary for each identified file~\ref{fig:demo:summarizer}.
Thus, we got all the information provided by upstream agents.

\begin{figure*}[h]
    \centering
    {\includegraphics[width=0.9\linewidth]{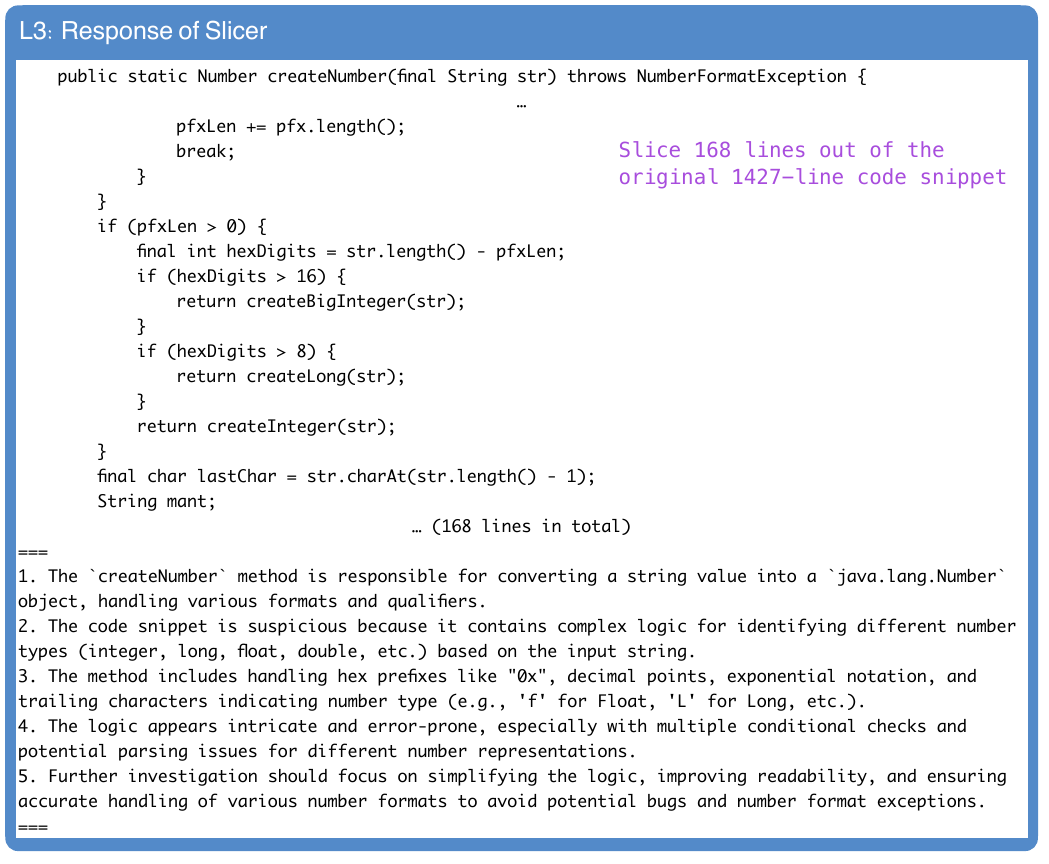}}
    \vspace{-0.1in}
    \caption{Response of \slicer.}
    \label{fig:demo:slicer}
\end{figure*}

\begin{figure*}[h]
    \centering
    {\includegraphics[width=0.9\linewidth]{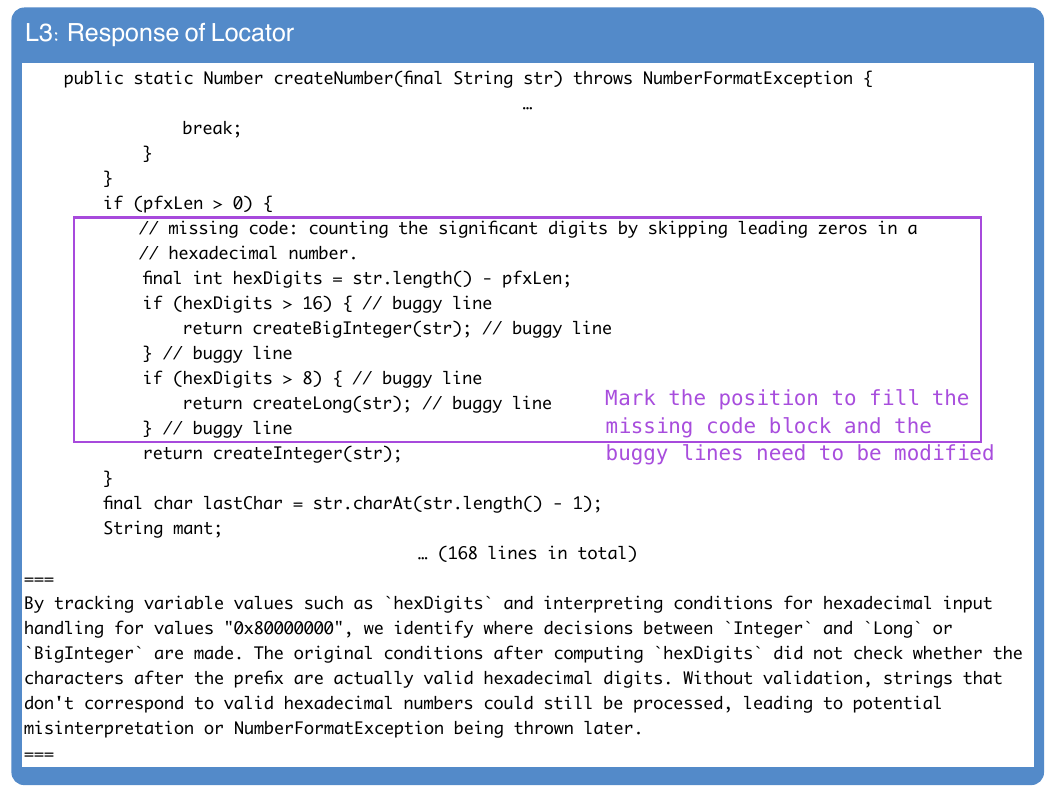}}
    \vspace{-0.1in}
    \caption{Response of \locator.}
    \label{fig:demo:locator}
\end{figure*}

\slicer extracts 168 suspicious code lines from 1427 lines in the original buggy code, largely narrowing down the examination scope, as shown in Figure~\ref{fig:demo:slicer}. 
\locator successfully pinpoints the root causes of this bug from the code lines sliced out~\ref{fig:demo:locator}.
The following agents can focus on the single logic conditions.

\begin{figure*}[h]
    \centering
    {\includegraphics[width=0.9\linewidth]{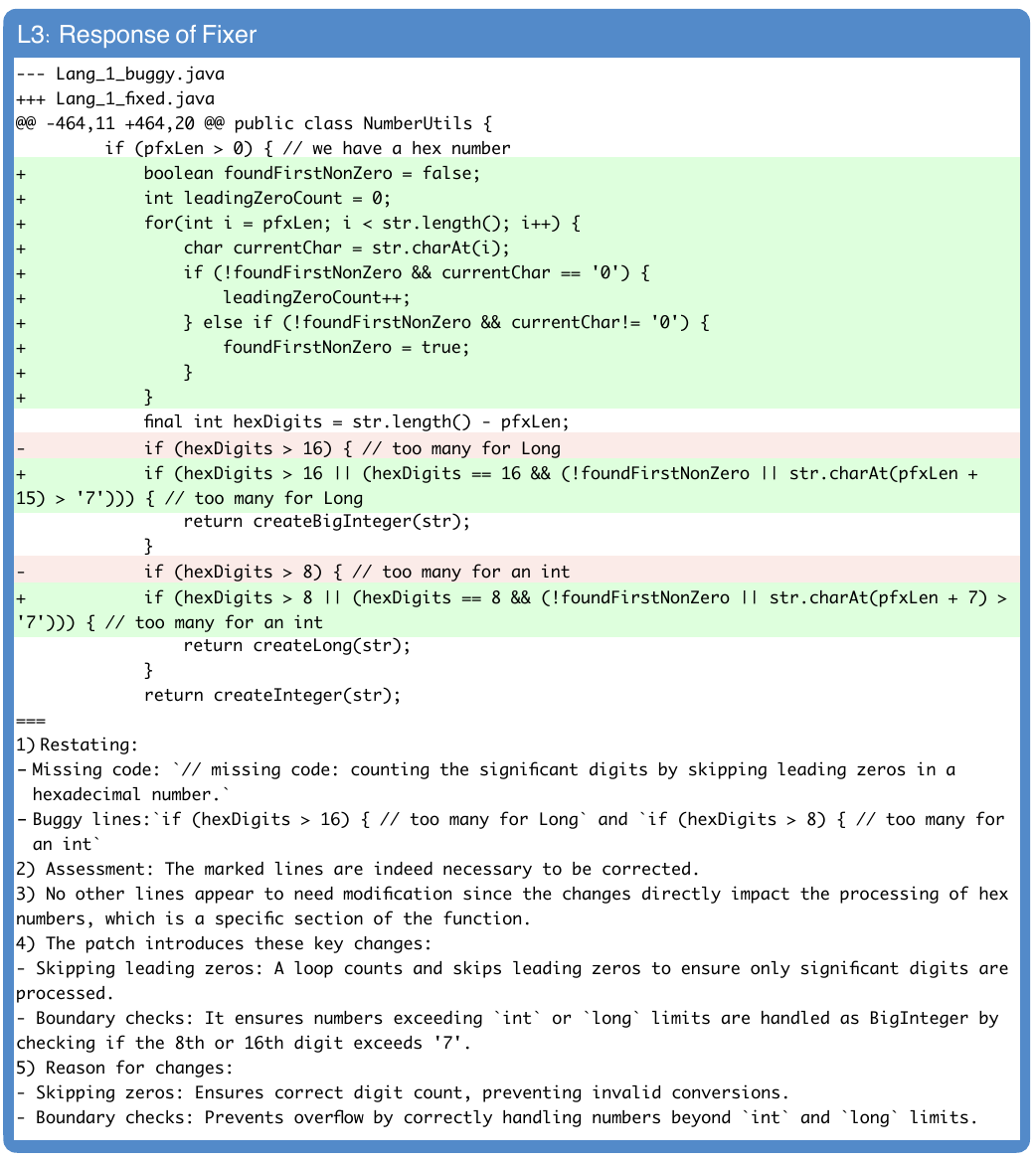}}
    \vspace{-0.1in}
    \caption{Response of \fixer.}
    \label{fig:demo:fixer}
\end{figure*}

\begin{figure*}[h]
    \centering
    {\includegraphics[width=0.9\linewidth]{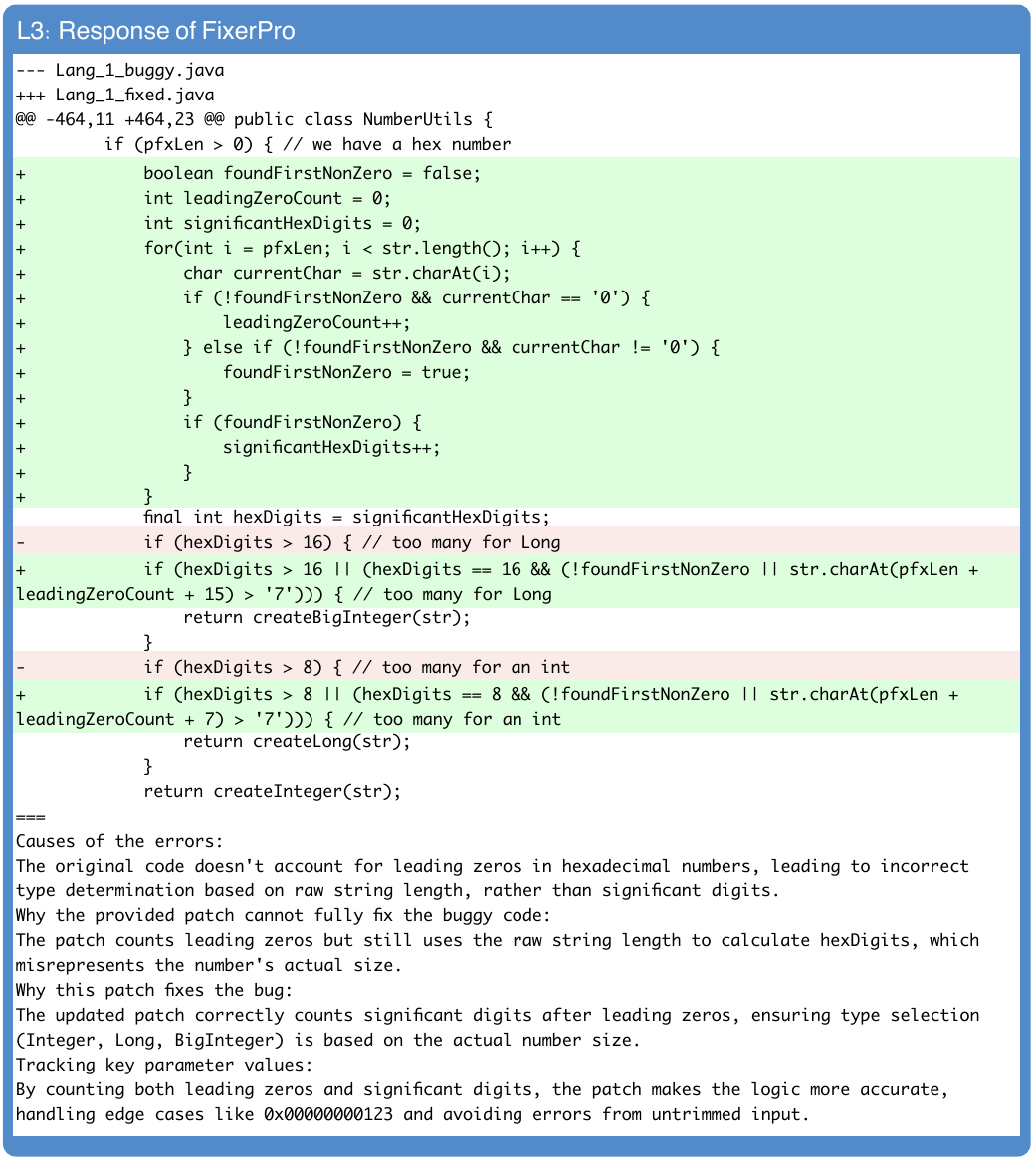}}
    \vspace{-0.1in}
    \caption{Response of \fixerpro.}
    \label{fig:demo:fixerpro}
\end{figure*}

However, \fixer failed to generate a plausible patch, as displayed in Figure~\ref{fig:demo:fixer}. It attempts to fulfill the missing code block by counting the valid digits of the hexadecimal number but causes incorrect type determination when a large number of leading zeros are present in a hexadecimal string. The patched code also assumes the hex digits should be directly compared based on raw string length without adjusting for those leading zeros.
\fixerpro identified the causes of errors made by \fixer, and provides an optimized patch. 
The fixed version, presented in Figure~\ref{fig:demo:fixerpro}, properly calculates the significant digits by counting the non-zero characters after the prefix and leading zeros. It also adjusts the comparisons for handling 16-digit and 8-digit boundary checks, ensuring that only significant digits are considered when deciding if the value is too large for ``Integer'' or ``Long''.

%% file: tables/algo.tex
\begin{algorithm}[htbp]
    \small
    \caption{\tool} \label{algo}
    \LinesNumbered
    \KwIn{
    $k$: number of maximum debugging attempts;
    $m$: number of maximum re-sampling attempts of a single agent;
    $bug\_meta$: metadata of the bug, a directory including code from the bug-located file, failing test case(s), errors, and program requirements.
    }
    \KwOut{patch, analysis}
    
    \SetKwFunction{FLone}{L1Repair}
    \SetKwFunction{FLtwo}{L2Repair}
    \SetKwFunction{FLthree}{L3Repair}
    \SetKwFunction{FDeg}{Debugging}
    \SetKwProg{Fn}{Function}{:}{End}
    
    \Fn{\FLone{$m$, bug\_meta, extra\_info}}{
        \For{$j \leftarrow 1$ to $m$}{
            marked\_code $\leftarrow$ Locator(bug\_meta, extra\_info) \\
            \If{ValidMarks(marked\_code)}{
            bug\_meta[code] $\leftarrow$ marked\_code \\
            break}
        }
        patch $\leftarrow$ Fixer(bug\_meta, extra\_info)\\
        \Return{patch} 
    }

    \Fn{\FLtwo{$m$, bug\_meta, extra\_info}}{
        \If{summary \textbf{NOT IN} extra\_info}{
            summary $\leftarrow$ Summarizer(bug\_meta[code]) \\
            extra\_info $\leftarrow$ \textbf{concat[}extra\_info; summary\textbf{]}
        }

        \For{$j \leftarrow 1$ to $m$}{
            snippet $\leftarrow$ Slicer(bug\_meta) \\
            \If{ValidSnippet(snippet)}{
            bug\_meta[code] $\leftarrow$ snippet \\
            break \\
            }
        }
        patch $\leftarrow$ \FLone{$m$, bug\_meta, extra\_info} \\
        patch, analysis $\leftarrow$  FixerPro(patch, Testing(patch), bug\_meta, extra\_info) \\
        \Return{patch, analysis} 
    }

    \Fn{\FLthree{$m$, bug\_meta, extra\_info}}{
        references $\leftarrow$ Helper(bug\_meta) \\
        FileList $\leftarrow$ RepoFocus(bug\_meta) \\
        summary $\leftarrow$ ArrayList() \\
        \For{file in FileList}{
            summary.append(Summarizer(ReadFile(file))) \\
        }
        extra\_info $\leftarrow$ \textbf{concat[}extra\_info; references; summary\textbf{]} \\
        \Return{\FLtwo{$m$, bug\_meta, extra\_info}}
    }
    
    \Fn{\FDeg{$k$, $m$, bug\_meta}}{
    \For{$i \leftarrow 1$ to $k$}{
            extra\_info $\leftarrow$ EmptyList()
            
            patch $\leftarrow$ \FLone{$m$, bug\_meta, extra\_info} \\
            \If{Testing(patch)}{\Return{patch, EmptyString()}}

            patch, analysis $\leftarrow$ \FLtwo{$m$, bug\_meta, extra\_info} \\
            \If{Testing(patch)}{\Return{patch, analysis}}

            patch, analysis $\leftarrow$ \FLthree{$m$, bug\_meta, extra\_info} \\
            \If{Testing(patch)}{\Return{patch, analysis}}

            patch, analysis $\leftarrow$ RefineAgents($m$, bug\_meta, extra\_info, patch) \\
            \Return{patch, analysis}
            
        }
    }
    
\end{algorithm}